\documentclass[twocolumn,english,aps,reprint, superscriptaddress,showpacs,longbibliography,showkeys,prx]{revtex4-2}
\usepackage{amsmath,amssymb,bbm,dsfont, mathrsfs,bm,braket,color,graphicx,comment} 
\usepackage[table]{xcolor}
\usepackage{tabularray}
\usepackage[colorlinks,citecolor=blue,urlcolor=blue,linkcolor = blue]{hyperref}
\usepackage{soul}

\usepackage[mathscr]{euscript}

\usepackage{hyperref}

\usepackage{tikz}
\usetikzlibrary{fadings}

\usepackage{multirow}

\newcommand\scalemath[2]{\scalebox{#1}{\mbox{\ensuremath{\displaystyle #2}}}}
\newcommand{\mapsfrom}{\mathrel{\reflectbox{$\mapsto$}}}

\newcommand{\wi}[2][logical]{%
\varepsilon^{\mathrm{#1}}_{#2}
}

\newcommand{\dwl}[1]{%
\Delta^{\mathrm{weak}}_{\textrm{logical}\if\relax\detokenize{#1}\relax\else,\fi #1}
}
\newcommand{\pef}{c}

\newcommand{\danch}[1]{%
d_{\textrm{anchor}\if\relax\detokenize{#1}\relax\else,\fi #1}
}
\newcommand{\decifit}{deficit}

\usepackage{marginnote}

\definecolor{dblu}{rgb}{0.4117,0.5764,0.7882}

\usepackage{cases}

\usepackage[plain]{algorithm2e}
\usepackage{colortbl}

\begin{document}

\title{Quantum optimization with globally driven neutral atom arrays}

\author{Martin Lanthaler}
\email[These authors contributed equally.\\Corresponding author: ]{Martin.Lanthaler@uibk.ac.at}
\affiliation{Institute for Theoretical Physics, University of Innsbruck, A-6020 Innsbruck, Austria}
\affiliation{Parity Quantum Computing GmbH, A-6020 Innsbruck, Austria}
\author{Kilian Ender}
\email[These authors contributed equally.\\Corresponding author: ]{Martin.Lanthaler@uibk.ac.at}
\affiliation{Institute for Theoretical Physics, University of Innsbruck, A-6020 Innsbruck, Austria}
\author{Davit Khachatryan}
\email[These authors contributed equally.\\Corresponding author: ]{Martin.Lanthaler@uibk.ac.at}
\affiliation{Parity Quantum Computing GmbH, A-6020 Innsbruck, Austria}
\author{Pedro Ildefonso}
\affiliation{Institute for Theoretical Physics, University of Innsbruck, A-6020 Innsbruck, Austria}
\affiliation{Parity Quantum Computing GmbH, A-6020 Innsbruck, Austria}
\author{Andrew Byun}
\affiliation{Institute for Theoretical Physics, University of Innsbruck, A-6020 Innsbruck, Austria}
\author{Clemens Dlaska}
\affiliation{Institute for Theoretical Physics, University of Innsbruck, A-6020 Innsbruck, Austria}
\affiliation{Digital Cardiology Lab, University Clinic of Internal Medicine III, Medical University of Innsbruck, A-6020 Innsbruck, Austria}
\author{Michael Schuler}
\affiliation{Parity Quantum Computing GmbH, A-6020 Innsbruck, Austria}
\author{Wolfgang Lechner}
\affiliation{Institute for Theoretical Physics, University of Innsbruck, A-6020 Innsbruck, Austria}
\affiliation{Parity Quantum Computing GmbH, A-6020 Innsbruck, Austria}
\affiliation{Parity Quantum Computing Germany GmbH, 20095 Hamburg, Germany}
\affiliation{Parity Quantum Computing France SAS, 75016 Paris, France}

\begin{abstract}
Neutral atoms trapped in configurable tweezer arrays are a promising platform for solving hard optimization problems. 
While such platforms natively embed unit-disk maximum weight independent set (\textsf{UD-MWIS}) problems, general problem classes require an embedding that maps them onto \textsf{UD-MWIS} instances. 
Implementing the necessary asymmetric weights has so far required local control of detunings at each atom. 
Here, we demonstrate quantum optimization using strictly global driving fields, eliminating the need for local field control. 
Instead, to effectively implement the required local detunings we exploit weak Rydberg interactions---too weak to generate Rydberg blockade but strong enough to effectively induce the required local detunings. 
For this, we introduce the concept of precisely placed \textit{anchor} atoms that impose controlled energy shifts on nearby atoms which can be effectively considered as induced local detunings.
We demonstrate that parity-architecture-based neutral atom embeddings of general optimization problems allow for a successful anchor placement strategy, such that all demanded local fields can be replaced by anchors.
We experimentally demonstrate this concept on a \textit{Pasqal Orion Alpha} machine and benchmark all standard building blocks.
Finally, we solve quadratic unconstrained binary optimization (QUBO) problems on four-node all-to-all connected graphs embedded in a 43-atom array using seven precisely-placed anchor atoms.
By removing the requirement for local addressing, our work overcomes a major experimental bottleneck in solving optimization problems on neutral-atom quantum hardware and can be readily integrated into existing platforms.
\end{abstract}

\maketitle

\section{Introduction}
\label{sec:Introduction}

\begin{figure*}[t]
\begin{centering}
\includegraphics[width = \textwidth]{Figure1.pdf}
\par\end{centering}
\protect\caption{\textit{Overview.}
(a) Concept of replacing site-dependent detunings $\Delta_i$ with effective detunings $\Delta_i^{\rm weak}$ induced by precisely placed, weakly interacting ``anchor'' atoms.
This is accomplished by an excited anchor atom inducing a distance-dependent energy contribution from vdW interaction as shown in (b).
Excited port (anchor) atoms are denoted with red-filled squares (circles). 
Note that in the inset the energy shift from the excitation of the anchor is ignored for visual clarity.
(c) A combinatorial optimization problem (OP) is mapped onto neutral-atom hardware using a parity transformation followed by an embedding with Rydberg gadgets, such as \textsf{LINK} and \textsf{3BODY} gadgets.
The fluorescence image shows an initial defect free array with atoms initialized in $\ket{0}$.
The variables of the parity-encoded problem are mapped to port atoms or \textsf{LINKs}, as indicated by colors. The corresponding weights of the original \textsf{UD-MWIS} problem would usually be implemented via site-dependent detunings at the ports or \textsf{LINKs}.
We augment this layout with auxiliary anchor atoms, whose spatial positioning allows us to encode the OP with a global detuning instead of site-dependent detunings.
(d) The solution to the OP can be found using a piecewise linear adiabatic annealing protocol (sweeping the global detuning from $\Delta_0 <0$ to $\Delta > 0$).
(e) Final state of the system after measuring.
Therein, circles denote the original positions of atoms now in the Rydberg state.
The shown configuration is the ground state for the given layout and solves the original OP.
}
\label{fig:Fig1}
\end{figure*}

Neutral atoms are a pioneering platform for both quantum simulation~\cite{Browaeys2020, Scholl_2021, Ebadi2021, daley2022, king2024computationalsupremacyquantumsimulation} and quantum computing~\cite{Saffman2010, Henriet2020,Bluvstein2022, Bluvstein2024, Bluvstein2025}, enabled by optical tweezer arrays which allow for arbitrary arrangements of thousands of atoms~\cite{Barredo2018,Kaufman_2021, Manetsch2024}. 
One key application for analog neutral atom quantum annealers is quantum optimization~\cite{Kim2022, Ebadi2022}, where combinatorial optimization problems are solved on quantum devices. 
Such optimization problems are ubiquitous in many fields of science and industry with many of them being \textsf{NP}-hard~\cite{Paschos2014, Lucas2014, Wurtz2024MIS}. 
Since classical computers struggle to solve these efficiently, such problems offer a fertile ground for exploring quantum speedups and near-term quantum utility~\cite{Henriet2020, Ebadi2022}.
One particular type of optimization problems, namely, maximum (weight) independent set (\textsf{M}(\textsf{W})\textsf{IS}) problems on unit-disk (\textsf{UD}) graphs, can be naturally encoded in the ground state of neutral-atom arrays~\cite{Pichler2018, pichler2018computationalcomplexityrydbergblockade,Kim2022, Ebadi2022} using the Rydberg blockade mechanism~\cite{Jacksch2000,Lukin2001,Gaetan2009,Urban09}.
First large-scale experiments on \textsf{UD-MIS} instances on so-called King's graphs~\cite{Ebadi2022} have demonstrated the effectiveness of Rydberg quantum annealers in solving such problems on hundreds of qubits and claimed a potential quantum scaling advantage. However, this claim was later on disproven and it was shown that very large instances of this very specific problem class can be solved very efficiently on classical computers~\cite{Andrist2023,Schuetz2025}.

To overcome this limitation and further explore the potential for quantum utility in neutral-atom-based quantum optimization, it is imperative to extend the capabilities of Rydberg quantum annealers to a broader and more complex class of optimization problems, such as quadratic or higher-order unconstrained binary optimization (QUBO/HUBO). 
This transition requires mapping general optimization problems on arbitrary graphs onto \textsf{UD-MWIS} instances. 
Such a mapping can be engineered by introducing specifically designed layouts of auxiliary atoms, or Rydberg gadgets~\cite{Kim2022, byun2024qubo, Nguyen2023, Lanthaler2023Ryblo, Stastny2023}, implemented through the Parity encoding~\cite{Lanthaler2023Ryblo}, crossing lattices~\cite{Nguyen2023}, or quantum wires~\cite{Kim2022}. The resulting atomic layouts encode the solution of optimization problems in their ground state, providing a viable pathway toward physical implementation of QUBO/HUBO problems in Rydberg quantum annealers.
However, the Rydberg gadgets typically rely on site-dependent atom detunings implemented via tightly focused laser beams~\cite{Labuhn2014, deOliveira2024, Kombe2025}, which remains experimentally demanding, and demonstrations of optimization with locally controlled neutral-atom arrays are still limited~\cite{deOliveira2024, Kombe2025}.
Although several approaches have been proposed to avoid this requirement, they are closer to heuristic designs of atomic arrays~\cite{byun2024qubo, Park2024FactoringRydberg} than a general approach.
Therefore, developing programmable primitives based on globally driven fields~\cite{Cesa2023, hu2025}, rather than relying on site-dependent addressing, is highly desirable, as it can reduce experimental demands and simplify scalability to larger systems.

Here, we close this gap and demonstrate quantum optimization of binary optimization problems on globally driven neutral atom arrays by utilizing \textit{weak interactions} between Rydberg atoms~\cite{Jacksch2000, Jo2020, Zhao2023, Mitra2023, Kim2023, Ming2025}.
We introduce additional, precisely placed atoms---which we call \textit{anchors}---that weakly interact with  target atoms. This weak interaction is not strong enough to induce Rydberg blockade, but strong enough to induce an energy shift on the target atom(s) mimicking the required local detuning [see Fig.~\ref{fig:Fig1} for an overview of the concept].
We showcase explicitly how this concept of utilizing weak interactions can be exploited to embed arbitrary combinatorial optimization problems on globally driven Rydberg quantum annealers by using the Parity framework~\cite{Lanthaler2023Ryblo, Lechner2015, Ender2023}. 
We experimentally benchmark our approach on a Pasqal Orion Alpha/Fresnel machine~\cite{darcangelo2024, Leclerc2025} by a detailed analysis of all necessary building blocks.
Finally, we demonstrate quantum optimization of small, all-to-all connected QUBO problems requiring 43 atoms.
This includes seven anchor atoms that encode the coefficients of the optimization problem via their positions rather than relying on local detunings.

The rest of this paper is organized as follows:
Section~\ref{sec:HamiltonianEncoding} reviews quantum optimization with Rydberg atom-based quantum annealers; 
Sec.~\ref{sec:weak_interactions} introduces the key mechanism to induce effective local detunings through weak interactions provided by additionally placed anchor atoms; 
Sec.~\ref{sec:parity_rydberg_encoding} demonstrates how this concept can be exploited to solve arbitrary combinatorial optimization problems on globally driven Rydberg quantum annealers using the Parity framework;
Sec.~\ref{sec:exps} experimentally benchmarks all necessary Rydberg gadget building blocks of the Parity architecture and demonstrates the solution of small all-to-all connected QUBO problems on a globally driven Rydberg quantum annealer; 
Finally, Sec.~\ref{sec:conclusion} provides outlooks and concludes our paper.

\section{Solving optimization problems in Rydberg Quantum Annealers}
\label{sec:HamiltonianEncoding}

In Rydberg quantum annealers, neutral atoms are usually trapped in optical tweezers with the ability to arrange them in nearly arbitrary configurations in two or even three dimensions \cite{Barredo2018}.
Each atom realizes a qubit where the electronic ground state represents the state $\ket{0}$ and a highly excited Rydberg state represents the state $\ket{1}$. 
These states can be coherently coupled via laser light resulting in an effective model described by the Hamiltonian
\begin{equation}
    \label{eq:RydbergHamiltonian}
    H_{\mathrm{Ryd}}(t) =  H_{\mathrm{diag}}(t) + H_{\mathrm{drive}}(t)
\end{equation}
with the diagonal part
\begin{equation}
    \label{eq:RydbergClassical}
    H_{\mathrm{diag}}(t) =  - \Delta(t) \sum_i  \hat n_i+ \sum_{i<j}\frac{C_6}{|\boldsymbol{x}_i-\boldsymbol{x}_j|^6}\hat n_i \hat n_j \, ,
\end{equation}
and the driver Hamiltonian 
\begin{equation}
    \label{eq:RydbergDriver}
    H_{\mathrm{drive}}(t) =  \frac{\Omega(t) }{2}\sum_i \hat\sigma_i^x \, .
\end{equation}
Here, $\Omega(t)$ denotes the Rabi-frequency, i.e., the strength of the laser coupling $\hat\sigma_i^x={\ket{0}}\!{\bra{1}}_i+{\ket{1}}\!{\bra{0}}_i$ acting on atom $i$ at position $\boldsymbol{x}_i$.
Furthermore, $\Delta(t)$ denotes the corresponding laser detuning, and
${\hat n_i= {|1\rangle}\!{\langle 1|}}_i$ is the projector onto states having a Rydberg excitation on the $i$-th atom.
The second term in $H_{\mathrm{diag}}$ describes the van der Waals (vdW) interaction between two Rydberg-excited atoms, with interaction coefficient $C_6>0$, giving rise to the Rydberg blockade mechanism~\cite{Jacksch2000,Lukin2001,Saffman2010}.
While in general both the Rabi frequency $\Omega(t)$ and detuning $\Delta(t)$ can be site dependent, in this work we assume only global addressing, which strongly reduces experimental complexity.

The Rydberg blockade energetically prevents more than one atom within the blockade radius from being excited to its Rydberg state. This mechanism provides a native, overhead-free way to encode and solve maximum independent set (\textsf{MIS}) problems on unit-disk (\textsf{UD}) graphs ${G=(V,E)}$~\cite{Pichler2018}. 
More precisely, \textsf{UD-MIS} is the problem of finding the largest subset of vertices $I \subseteq V$ that are not connected by an edge $e\in E$~\cite{Garey1983NPcompleteness} on graphs constructable from a set of 2D points with edges whenever two points are within unit distance~\cite{Clark90UnitDiscGraphs}.
To see the connection to the Rydberg Hamiltonian, consider only the diagonal part $H_{\mathrm{diag}}$ of $H_{\rm Ryd}$ and assume a constant $\Delta > 0$.
In this regime, the Rydberg blockade (energetically) forbids more than one Rydberg excitation on atoms within a characteristic distance called \textit{(static) blockade radius} ${r_B = \left(C_6/\Delta\right)^{1/6}}$.
Identifying Rydberg excited atoms as being in the independent set $I$ allows one to directly encode the independent set condition by defining an edge of the \textsf{MIS} problem to each pair of atoms if and only if they lie within $r_B$.
With this encoding, the \textsf{UD-MIS} problem can be solved by finding the ground state of the corresponding Rydberg Hamiltonian, where atoms are placed such that Rydberg blockade conditions match the target unit-disk graph.
In Rydberg quantum annealers the search for the ground state can be achieved through quantum annealing based approaches or variational quantum algorithms~\cite{Ebadi2022}. 

In order to embed arbitrary graphs beyond \textsf{UD} or more general optimization problems, an additional encoding has to be applied.
Common encodings such as the crossing lattice~\cite{Nguyen2023}, 3D quantum wires~\cite{Kim2022} or Parity-based approaches~\cite{Lanthaler2023Ryblo} rely on so-called Rydberg gadgets.
Those are frequently appearing building blocks consisting of a few specifically arranged atoms, which are precisely designed to realize target binary logical relations among a subset of the atoms through the Rydberg blockade mechanism.
The amalgamation of such gadgets~\cite{Stastny2023} to larger atomic layouts then allows to encode general optimization problems into Rydberg quantum annealers [see Fig.~\ref{fig:Fig1}(c) for an illustration]. 
However, all these approaches require to set specific local detunings on individual atoms, since they encode the target logical operations as a weighted \textsf{UD-MIS} instance (\textsf{UD-MWIS}). 

In contrast to the unweighted setting, \textsf{MWIS} problems are defined on a weighted graph $(G,w:V\rightarrow \mathbb{R})$ and consist of finding an independent set $I$ that maximizes the total weight ${W(I) = \sum_{i\in I}w_i}$.
The \textsf{MWIS} problem can be formulated as a ground-state search task for the Hamiltonian
\begin{equation}
H_{\mathsf{MWIS}} = - \sum_{i\in V} w_i \hat n_i + U\sum_{(i, j)\in E}\hat n_i\hat n_j \, ,
\label{eq:MWISHamiltonian}
\end{equation}
where the first term maximizes the total weight and the second term enforces the independent set condition.
In particular, iff $0<w_i < U$ is satisfied, the ground state
of $H_{\mathsf{MWIS}}$ faithfully encodes a solution of the \textsf{MWIS} problem.
Similar to the unweighted \textsf{UD-MIS} problem discussed above, \textsf{UD-MWIS} problems can be natively embedded into Rydberg annealers by enforcing independece via Rydberg blockade [cf.\ Eq.~\eqref{eq:RydbergClassical} and Eq.~\eqref{eq:MWISHamiltonian}]. 
However, encoding the weights $w_i$ typically requires local detunings on individual atoms, $\Delta \sum_i \hat{n}_i \rightarrow \sum_i \Delta_i \hat{n}_i$ with $\Delta_i \propto w_i$.
To avoid this complexity, in the following sections we develop a toolkit that allows us to encode---and ultimately solve---arbitrary optimization problems on globally driven neutral atom arrays.
To this end, the encoding is fully described by the Hamiltonian in Eq.~\eqref{eq:RydbergClassical}, which significantly reduces the experimental hardware requirements.

\section{Effective local detunings from weak interactions}
\label{sec:weak_interactions}

\begin{figure}[t]
\begin{centering}
\includegraphics[width =\columnwidth]{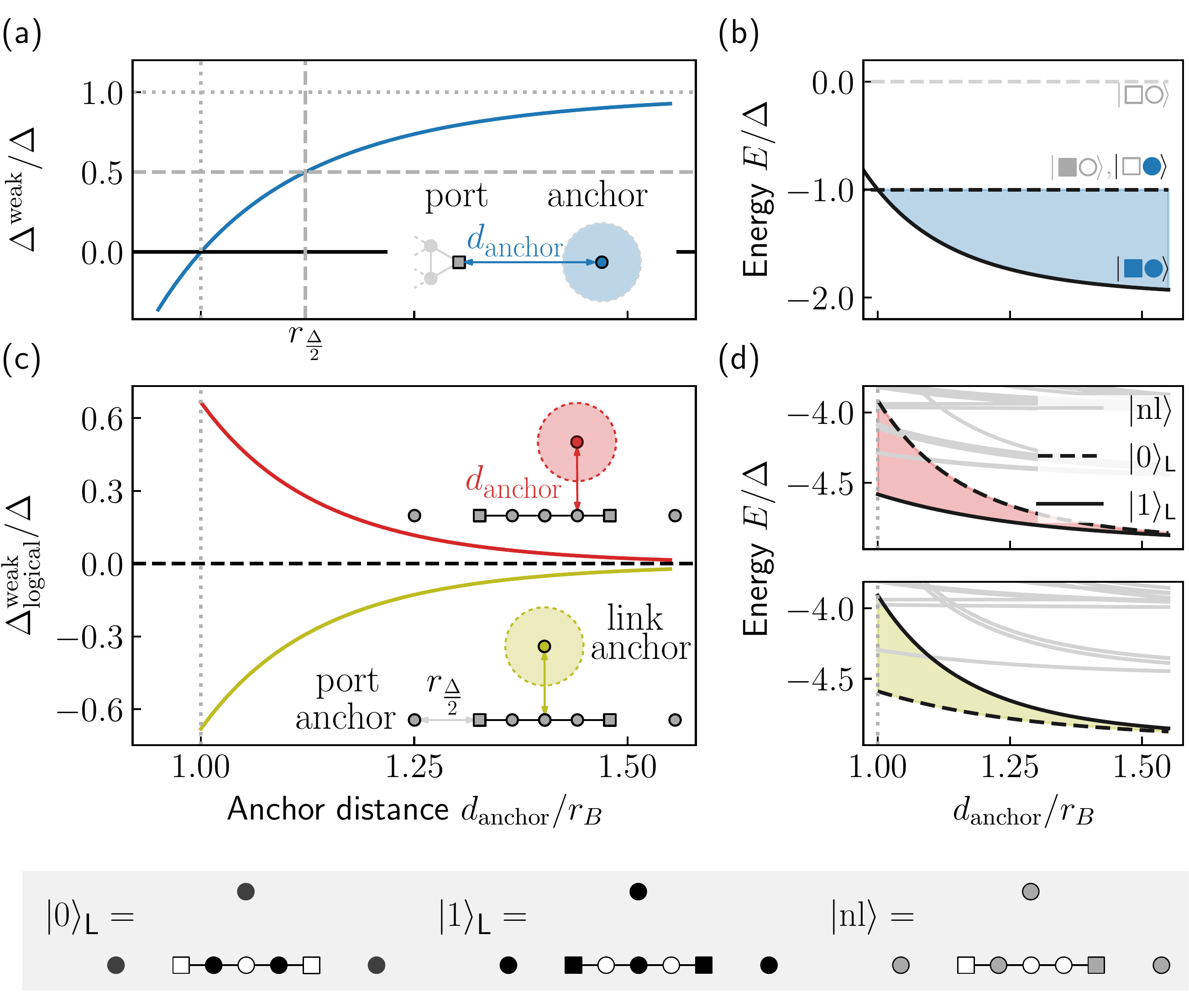}
\par\end{centering}
\protect\caption{\textit{Placing anchors}. 
(a) Effective detuning $\Delta^{\rm weak}$ on the port atom induced through vdW interaction with the anchor atom, as a function of their distance $d_{\rm anchor}$. 
(b) Energy spectrum of the port-anchor system as a function of $d_{\rm anchor}$.
The blue-shaded region shows the energy difference (corresponding to $\Delta^{\rm weak}$) between the states where the anchor is excited in the weak-interaction regime, $d_{\rm anchor} > r_B$.
(c) Effective detuning $\dwl{}$ induced by the link anchor on the \textsf{LINK} gadget as a function of the link anchor distance $d_{\rm anchor}$, which is positive (negative) for an offset (center) link anchor. 
Port anchors induce the required local detunings on the ports (see main text).
(d) Low-energy spectra of the full atomic system. 
The energy difference between the two logical states $\ket{0}_\textsf{L}$ and $\ket{1}_\textsf{L}$ (see legend) defines $\dwl{}$ and is highlighted in color. Non-logical states are shown in gray. 
The legend depicts the logical states and an example of a non-logical state.
In all figures, port atoms are depicted as squares and excited atoms are shown with filled symbols.
}
\label{fig:weak_interaktion}
\end{figure}

In this section, we propose using weak vdW interactions from additional, precisely placed auxiliary atoms---which we term  \textit{anchors}---to induce local detuning shifts on nearby atoms [see Fig.~\ref{fig:Fig1}(a)]. 
These anchors are positioned outside the blockade radius of any atom within the original atomic layout, ensuring they are only weakly interacting~\cite{Jacksch2000, Jo2020, Zhao2023, Mitra2023, Kim2023, Ming2025}.
As detailed below, in this regime, they induce state-dependent energy shifts that act as effective local detunings while preserving the Rydberg blockade conditions within the original atomic layout.
Consequently, we will show that the local detunings $\Delta_i$ required to encode optimization problems via the mapping to Eq.~\eqref{eq:MWISHamiltonian} can be emulated in globally driven Rydberg quantum annealers through appropriately placed anchors [see also Sec.~\ref{sec:parity_rydberg_encoding}].

To explain the core idea, let us start with the simplest atomic layout consisting only of a single atom $i$ with a global detuning $\Delta > 0$ and $\Omega \rightarrow 0$. 
We place an additional anchor atom at a distance $d_{\mathrm{anchor}, i} > r_B$ which weakly interacts with atom $i$ and is subject to the same global detuning [cf.\ inset in Fig.~\ref{fig:weak_interaktion}(a)].
This system is described by the Hamiltonian
\begin{equation}
\label{eq:H_anchor}
    H_{\mathrm{anchor}, i} = -\Delta \left(\hat{n}_i  + \hat{n}_{\mathrm{anchor}}\right) +\wi[]{i} \hat{n}_{i}\hat{n}_{\mathrm{anchor}} \, ,
\end{equation}
where the interaction strength is given by
\begin{equation}
\label{eq:anchor_delta}
    \wi[]{i} = \frac{C_6}{\danch{i}^6} \, .
\end{equation}
In the weakly interacting regime $0 < \wi[]{i} < \Delta$ the anchor does not induce blockade on atom $i$, such that the low-energy subspace of the atoms is spanned by configurations where the anchor atom is excited to the Rydberg state $\ket{1}_{\rm anchor}$, as shown in  Fig.~\ref{fig:weak_interaktion}(b)~\footnote{Please note, that in this basic example atom $i$ and the anchor are equivalent by symmetry. 
In light of our goal to place anchors next to atoms that are part of a larger layout we here identify the states with an excited anchor as the relevant subspace.
}.
In this regime Eq.~\eqref{eq:H_anchor} transforms into
\begin{equation}
    H_{\mathrm{anchor}, i} \xrightarrow[\hat{n}_\mathrm{anchor} \rightarrow 1]{} -(\Delta - \wi[]{i}) \hat{n}_i  + {\rm const.}\, ,
\end{equation}
such that the anchor effectively changes the detuning locally on atom $i$ to 
\begin{equation}
    \Delta_i^{\mathrm{weak}}(\danch{i}) = \Delta - \wi[]{i}(\danch{i}).
    \label{eq:weak_programming}
\end{equation}
Changing the distance $\danch{i}$ between the anchor atom and atom $i$ directly modifies this effective detuning, as shown in Fig.~\ref{fig:weak_interaktion}(a).%
This tunability allows to replace a demanded local detuning $\Delta_i$ with the induced detuning $\Delta_i^{\rm weak}$ by assigning
\begin{equation}
     \Delta_i \mapsfrom \Delta_i^\mathrm{weak}(\danch{i}).
     \label{eq:localdetuing_byweakinteraction_openport}
\end{equation}
Note that the weak interaction can only reduce the global detuning and that the effective local detuning is limited to the range $0<\Delta_i^\mathrm{weak}<\Delta$ when the anchor is placed far enough to not induce blockade, ${\danch{i} > r_B}$ [see Fig.~\ref{fig:weak_interaktion}(a)].
In practice, this constraint does not hinder our approach. 
The Rydberg gadgets considered here naturally align with this requirement [see Sec.~\ref{sec:parity_rydberg_encoding}], and the global energy scales can always be rescaled accordingly.

In the above example we have only considered the weak interaction between a pair of atoms. 
However, the general setting we consider consists of a multi-atom layout formed by amalgamating several Rydberg gadgets, as shown in Fig.~\ref{fig:Fig1}(c). 
These gadgets consist of two kinds of atoms, interior atoms and specific ``boundary'' atoms called \textit{ports}.
While the interior atoms are auxiliary atoms to create the gadgets' target functionality, the state of the port atoms typically encodes variables of the original optimization problem and their corresponding weights.
Furthermore, the Rydberg gadgets can be \textit{homogenized} such that local detunings are required only on port atoms and all interior atoms share a constant detuning, while preserving the original atomic layout [cf.\ Sec.~\ref{sec:parity_rydberg_encoding} and App.~\ref{app:Homogenization}]. 
In general layouts two cases appear: either ports are \textit{open}, i.e., not connected to other gadgets, or they are connected to other gadgets via \textsf{LINK}s.
In the following, we discuss an anchor placement strategy for both cases.

\paragraph*{(a) Open ports:}
First, we consider an open port, which is defined as a port not connected to another gadget. In this scenario, the port atom is typically geometrically well-isolated. This isolation allows an anchor to be placed such that it predominantly induces a local detuning on the port atom itself; due to the rapid $R^{-6}$ decay of the vdW interactions, its effect on the rest of the atomic layout remains minimal [see Fig.~\ref{fig:Fig1}(c)]. 
Then, Eq.~\eqref{eq:weak_programming} can be used to induce the desired local detuning on the port atom.
Treating the port interaction in isolation provides a first-order approximation that is often sufficient. 
However, higher-order corrections, accounting for long-range interactions with the remaining gadget atoms (and potentially adjacent \textsf{LINK} gadgets), can be incorporated to fine-tune the anchor positions for optimal precision.

\paragraph*{(b) \textsf{LINK}s:}
Gadgets are connected at their ports via \textsf{LINKs}---specialized atomic chains of an odd number of pairwise-blockaded atoms that enforce identical states at their start and end points [see Fig.~\ref{fig:Fig1}(c)], thereby acting as a copy gadgets. 
Crucially, each \textsf{LINK} also delocalizes the state encoded at the ports throughout its entire length allowing us to place an anchor---which we term a link anchor---near the center of the \textsf{LINK} rather than at the ports themselves. This avoids a fundamental issue: unlike open ports, connected ports are not spatially isolated and placing anchors directly there would induce detrimental non-perturbative effects on surrounding atoms.

We now demonstrate that the positioning of these link anchors safely induces the required port detunings non-locally. 
\textsf{LINK} gadgets are designed to have the states
\begin{equation}
|0\rangle_\textsf{L} := |010\dots 10\rangle \quad \text{and} \quad
|1\rangle_{\textsf{L}} := |101\dots 01\rangle,
\end{equation}
as their degenerate ground states. 
These states encode a single logical variable $\hat{n}_\textsf{L} := |1\rangle\!\langle 1|_\textsf{L}$  non-locally on the \textsf{LINK}.
Placing a link anchor will induce weak interactions 
\begin{equation}
\wi[]{i} = \frac{C_6}{d_i^6},
\end{equation}
on all the excited atoms of the \textsf{LINK}:
\begin{equation}
\tikz{
\draw[red!50!white,  line width=.4mm] (2 * .5,0) -- (2 * 1, 2 * .55);
\draw[red!60!white, line width=.4mm] (2 * 1.5,0) -- (2 * 1,2 * .55);
\draw[red!70!white, line width=.5mm] (2 * 1,0) -- (2 * 1, 2* .55);
\draw[red!30!white,  line width=.3mm] (0,0) -- (2 * 1, 2 * .55);
\draw[red!30!white,  line width=.3mm] (2 * 2,0) -- (2 * 1, 2* .55);
\draw[fill, black!50!white] (-2 * .58,-0.75) circle (.014 cm) node[left, black,  xshift=-.1cm, yshift=0.015cm,  font=\footnotesize]{$|1\rangle_{\mathsf{L}} = $};
\draw[fill, black!60!white] (-2 * 0.4,-0.75) circle (.014 cm);
\draw[fill, black!70!white] (-2 * 0.22,-0.75) circle (.014 cm);
%
\path (2,0) -- node[fill=white, yshift=-0.1cm,xshift=0.0cm]{$d_{0}$} (2, 1.2);
\path (3,0) -- node[fill=white, yshift=-0.1cm,xshift=0.07cm]{$d_{1}$} (2, 1.2);
\path (4,0) -- node[fill=white, yshift=-0.1cm,xshift=0.15cm]{$d_{2}$} (2, 1.2);
\draw[line width=0.3mm, black!90!white] (0,-0.75) -- (2 * 2,-0.75);
\draw[fill, draw=black] (0,-0.75) circle (.085 cm)  node[below, yshift=-.03cm, rotate=0]{$\wi[]{2}$};
\draw[fill, white, draw=black] (2 * 0.5,-0.75) circle (.085 cm); 
\draw[fill, draw=black] (2 * 1,-0.75) circle (.085 cm) node[below, yshift=-.03cm,  rotate=0]{$\wi[]{0}$}; 
\draw[fill, white, draw=black] (2 * 1.5,-0.75) circle (.085 cm); 
\draw[fill, draw=black] (2 * 2,-0.75) circle (.085 cm) node[below, yshift=-.03cm, rotate=0]{$\wi[]{2}$};
\draw[fill, red, draw=black] (2 * 1,2 * .55) circle (.085 cm) node[left, black,  xshift=-.2cm, yshift=0.045cm,  font=\footnotesize]{anchor};
\draw[fill, black!70!white] (2 * 2+ 2* 0.22,-0.75) circle (.014 cm);
\draw[fill, black!60!white] (2 * 2+ 2 * 0.4,-0.75) circle (.014 cm);
\draw[fill, black!50!white] (2 * 2+2 * .58,-0.75) circle (.014 cm);
%
%
\draw[fill, black!50!white] (-2 * .58,0) circle (.014 cm) node[left, black,  xshift=-.1cm, yshift=0.015cm,  font=\footnotesize]{$|0\rangle_{\mathsf{L}} = $};
\draw[fill, black!60!white] (-2 * 0.4,0) circle (.014 cm);
\draw[fill, black!70!white] (-2 * 0.22,0) circle (.014 cm);
\draw[line width=0.3mm, black!90!white] (0,0) -- (2 * 2,0);
\draw[fill, white, draw=black] (0,0) circle (.085 cm);
\draw[fill, black, draw=black] (2 * 0.5,0) circle (.085 cm) node[below, black,  yshift=-.03cm, rotate=0]{$\wi[]{1}$}; 
\draw[fill, white, draw=black] (2 * 1,0) circle (.085 cm); 
\draw[fill, black, draw=black] (2 * 1.5,0) circle (.085 cm)
node[below, black, yshift=-.03cm, rotate=0]{$\wi[]{1}$}; 
\draw[fill, white, draw=black] (2 * 2,0) circle (.085 cm);
\draw[fill, black!70!white] (2 * 2+ 2* 0.22,0) circle (.014 cm);
\draw[fill, black!60!white] (2 * 2+ 2 * 0.4,0) circle (.014 cm);
\draw[fill, black!50!white] (2 * 2+2 * .58,0) circle (.014 cm);
}
\label{eq:weak_link_be_strong}
\end{equation}
Here, $d_i$ denotes the distance of the anchor to the $i$-th neighbor of the center atom with an anchor distance $\danch{} = d_0$. 
Because of the different Rydberg excitation patterns, the anchor induces different energy shifts on the two logical states of the link, in particular
\begin{align}
\begin{split}
    \wi[]{\ket{0}_\textsf{L}} &= 2 \wi[]{1} + \dots,
    \\
    \wi[]{\ket{1}_\textsf{L}} &=\wi[]{0} + 2 \wi[]{2} + \dots\,.
    \label{eq:energy_shifts_link}
    \end{split}
\end{align}
Within the logical subspace $\{\ket{0}_\textsf{L}, \ket{1}_\textsf{L} \}$ the resulting energy difference can be described (up to constants) by an effective logical detuning Hamiltonian
\begin{align}
\label{eq:h_link_eff}
    H_{\textsf{LINK}, {\rm anchor}} &= -\dwl{}\hat{n}_{\textsf{L}}, \nonumber\\
    \dwl{}(\danch{}) &= \wi[]{\ket{0}_\textsf{L}} - \wi[]{\ket{1}_\textsf{L}} \ .
\end{align}
Note that this equation only describes the additional anchor-induced effect, and does not include the Hamiltonians for the bare anchor and \textsf{LINK} itself.
Equation~\eqref{eq:h_link_eff} shows that an anchor at the center of the link can effectively induce a local detuning for the logical operator $\hat{n}_{\textsf{L}}$ on the link, similar to the case of an open port.
In other words, the link anchor can effectively induce required local detunings at the connected ports non-locally along the \textsf{LINK}.

When the anchor is placed above the center atom, the largest energy shift $\varepsilon_0$ is applied to the $\ket{1}_\textsf{L}$ state because its center atom is in the Rydberg state.
Therefore, $0 < \wi[]{\ket{0}_\textsf{L}} < \wi[]{\ket{1}_\textsf{L}}$, and only negative logical detunings $\dwl{}<0$ can be induced.
However, in contrast to port anchors, both signs of detunings can be induced with link anchors.
To induce positive logical detunings, the link anchor can be moved to the atom left or right of the center atom [see offset anchor in Fig.~\ref{fig:weak_interaktion}(c)].
For such an offset anchor, the other logical state $\ket{0}_\textsf{L}$ experiences the largest energy shift, such that $\dwl{}> 0$.

Note that the above description holds for all \textsf{LINK} gadgets of length $4n+1$ (for integer $n\geq1$), since in these cases, the center atom shares the same state as the port atoms in the logical subspace. For \textsf{LINKs} of length $4n+3$, however, this relationship is reversed: the center atom is in the state $\ket{1}$ ($\ket{0}$) when the ports are in the state $\ket{0}$ ($\ket{1}$). The same methods still apply to these \textsf{LINKs}, with the only modification being that center anchors induce a positive $\dwl{}>0$, while offset anchors induce a negative $\dwl{}<0$.

In Fig.~\ref{fig:weak_interaktion}(c,d) we demonstrate both of the above introduced concepts, \textit{(a)} and \textit{(b)}, on an isolated \textsf{LINK} layout of length five. 
First, an isolated \textsf{LINK} requires local detuning of $\Delta/2$ on its two port atoms while all interior atoms experience the detuning $\Delta$.
This configuration makes the two logical states $\ket{0}_\textsf{L}$ and $\ket{1}_\textsf{L}$ the degenerate ground states.
Since both ports are \textit{open} for an isolated \textsf{LINK}, we can use method \textit{(a)} and place port anchors at a distance $r_{\frac{\Delta}{2}}$ from the left and right ports, as illustrated in Fig.~\ref{fig:weak_interaktion}(c).
This distance is set via Eq.~\eqref{eq:weak_programming} (with an added correction for the perturbative long-range vdW effects) such that the induced local detuning on the port atoms is $\Delta_i^{\rm weak} (r_{\frac{\Delta}{2}}) = \Delta/2$ for a global detuning $\Delta$.
Then, a logical detuning $\dwl{}$ on the link, splitting the energies between the two logical states, can be induced by placing a link anchor above the center, or next to center link atom, according to method \textit{(b)} described above [see colored atoms in Fig.~\ref{fig:weak_interaktion}(c)]. 

In Fig.~\ref{fig:weak_interaktion}(d) we plot the resulting low energy spectrum of the full link and anchors system under the Rydberg Hamiltonian Eq.~\eqref{eq:RydbergClassical} as a function of the distance $\danch{}$ between anchor and link.
We show both cases of a center link anchor (bottom) or an offset link anchor (top).
For large $\danch{} \gtrsim 1.5\, r_B$ the vdW interaction between link anchor and link atoms becomes negligible and the two target logical states $\ket{0}_\textsf{L}$ and $\ket{1}_\textsf{L}$ become the degenerate ground states. 
This demonstrates that the two port anchors are inducing the appropriate local detuning $\Delta_i^{\rm weak} = \Delta/2$ on the two port atoms.

Bringing the link anchors closer leads to an energy splitting between the logical states which can be interpreted as a programmable, anchor-induced, logical detuning on the \textsf{LINK} according to Eq.~\eqref{eq:h_link_eff}. 
The strength of the induced logical detuning $\dwl{}$ is plotted as a function of the anchor distance $\danch{}$ in Fig.~\ref{fig:weak_interaktion}(c).
As described above, a center (offset) link anchor yields a negative (positive) logical detuning. 
Also, we observe that its maximal strength is limited in this setting to $|\dwl{}| \lesssim 0.5 \Delta$ for $\danch{} > r_B$.
Finally, we want to note that non-logical states $\ket{\rm nl}$ of the link can become low-energy states even before $\danch{}$ reaches $r_B$, as observed in Fig.~\ref{fig:weak_interaktion}(d) (gray lines).
Such non-logical states do not correspond to \textsf{MWIS} solutions and they can limit the range of local detunings that can be effectively induced.
In principle, this does not limit our approach since global energy scales can be adjusted accordingly, but at the expense of potentially reduced energy gaps. 
The appearance and relevance of such non-logical states is non-trivial in larger layouts and in reality a trade-off has to be found between pushing them out from the low-energy sectors and reducing gaps among the logical \textsf{MWIS} states [see Sec.~\ref{sec:4body} for a discussion].

While the methods presented in this section do not allow for the engineering of completely arbitrary effective detuning patterns on arbitrary layouts with globally driven neutral atom arrays, such flexibility is ultimately not necessary for encoding generic optimization problems. 
As we will demonstrate in the following section Sec.~\ref{sec:parity_rydberg_encoding}, utilizing the parity encoding as an intermediate step translates the requirement for arbitrary detunings into a structured pattern---one that can be precisely engineered using weakly interacting anchors placed at open ports and \textsf{LINK}s.

\section{Global parity Rydberg-encoding}
\label{sec:parity_rydberg_encoding}

\begin{figure*}[th]
\begin{centering}
\includegraphics[width = \textwidth]{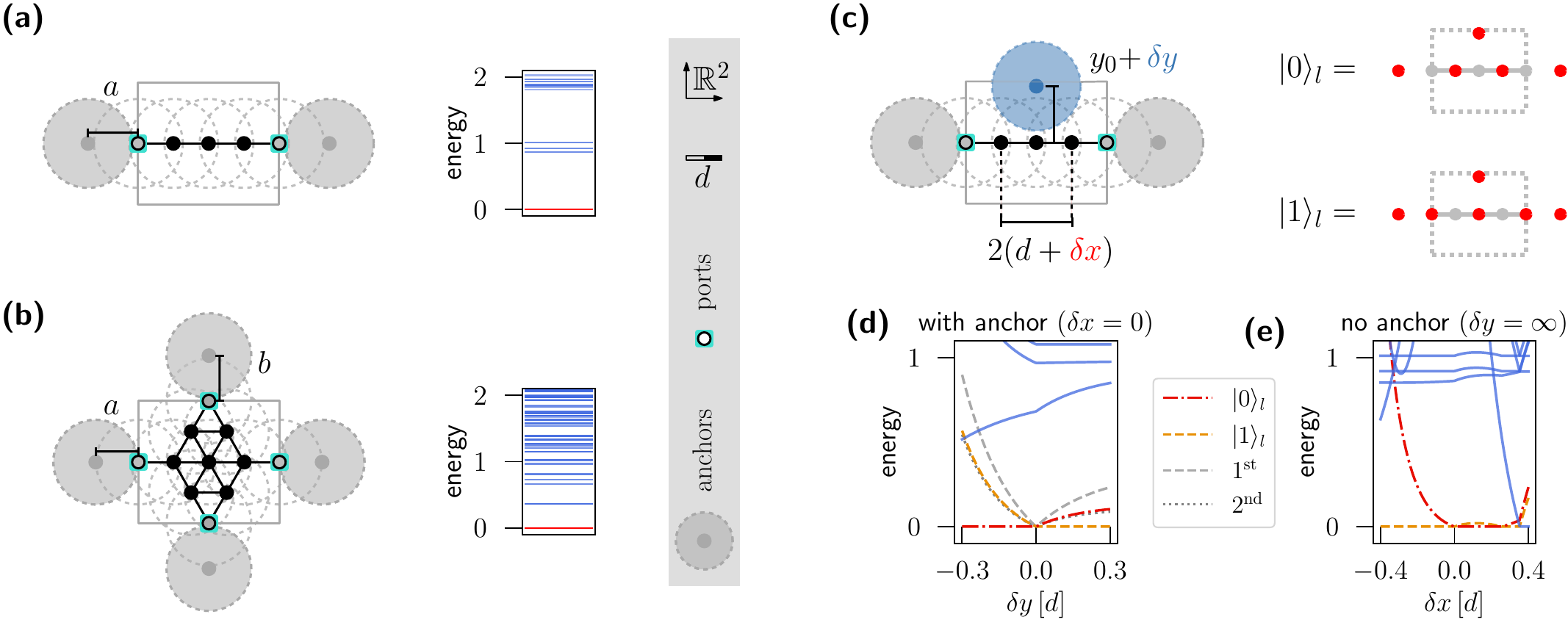}
\par\end{centering}
\protect\caption{\textit{Parity Rydberg layouts.} 
(a) Transformation of the standard parity layout (containing 3- and 4-body constraints) into an equivalent representation requiring 2- and 3-body constraints. This is achieved by breaking up 4-body constraints into two 3-body constraints. 
(b)
The resulting parity layout can be mapped onto 2D Rydberg quantum annealers by encoding 2-body constraints via \textsf{LINK} gadgets and 3-body constraints via homogenized \textsf{3BODY} gadgets.
\textsf{LINK}s are also added to separate other gadgets from each other to avoid crosstalk.
Additionally, we show frequently appearing structures resulting from amalgamation, such as \textsf{KITE} and \textsf{4BODY}.
In these representations, atoms are shown as dots and edges correspond to blockaded pairs of atoms.
Anchor atoms are highlighted in color together with their blockade radius. 
The color-coding corresponds directly to the local fields in the original parity layout shown in (a).
Numbers adjacent to ports denote the required gadget specific local port detunings $\Delta_{\mathrm{port}, e}$ (in units of $\Delta$) that are induced by the corresponding anchors.
Accordingly, black nodes represent a detuning of $\Delta$ and lighter colors indicate required detunings less than $\Delta$.
The dashed arrow in the center panel indicates, as an example, a ``shifting'' of detuning \decifit{}s from the lower port of the \textsf{KITE} gadget to the connected link, which can then be induced by the gray link anchor.
}
\label{fig:parity_layouts}
\end{figure*}

In Sec.~\ref{sec:weak_interactions}, we described how to use systematically placed anchor atoms to implement effective local detunings $\Delta_i^{\rm weak}$ and $\dwl{i}$, where $i$ labels either a port atom or a \textsf{LINK}.
Previous work~\cite{Lanthaler2023Ryblo} mapped combinatorial optimization problems (OPs) to an intermediate parity representation
and ultimately to a \textsf{UD-MWIS} formulation on a Rydberg atom array, but this implementation required locally addressable laser detunings.
In this section, by introducing anchors and incorporating newly developed homogenized gadgets (see App.~\ref{app:Homogenization}), we extend this pipeline to
\begin{equation}
    \text{OP} \mapsto \text{Parity} \mapsto \textsf{UD-MWIS} \mapsto \text{Global},
    \label{eq:overview_transformations}
\end{equation}
thus completely circumventing the need for local addressability.
We briefly review the first two steps that have been introduced in Ref.~\cite{Lanthaler2023Ryblo} below before detailing the final step $\textsf{UD-MWIS} \mapsto \text{Global}$.

First, an OP can be represented as the ground-state problem of a spin-glass Hamiltonian of the form
\begin{equation}
\scalemath{.94}{
H_{\mathrm{OP}}(\mathbf{s}) = \sum_{i\in E_1} J_i s_i + 
\!\!
\sum_{i,j\in E_2} J_{ij} s_i s_j + 
\!\!\! \sum_{i,j,k \in E_3} J_{ijk} s_i s_j s_k + \cdots,
}
\label{eq:spin_glass_OP}
\end{equation}
with real coefficients  ${J_{e}} \in \mathbb{R}$ defined on a (hyper-)graph with edges $e\in E_1 \cup E_2 \cup  E_3 \cup \cdots$. 
In the parity approach~\cite{Lechner2015, Ender2023}, rather than working directly with the spin variables $s_i$, parity variables encode relative information, i.e.,  $\sigma_e =\sigma_{i,j, \dots} = s_is_j\cdots\;$, where $e=(i,j,\dots)$ [step one of Eq.~\eqref{eq:overview_transformations}].
This transformation maps general spin-glass problems onto a connectivity-limited graph that is well suited for quantum hardware with short-range connectivity:
\begin{equation}
H_{\mathrm{parity}}(\boldsymbol{\sigma}) = \sum_{e} J_{e} \sigma_{e} - \lambda\sum_{k} C_k.
\label{eq:parity_hamiltonian}
\end{equation}
Here, the local field terms $J_e\sigma_e$ encode the initial OP and the geometrically local constraints $C_k$ ensure a consistent encoding in the low energy manifold (with $\lambda>0$)~\cite{Lechner2015, Ender2023}.
These constraints are defined as products of either three or four parity variables, $\sigma_{a}\sigma_{b}\sigma_{c}$ or $\sigma_{a}\sigma_{b}\sigma_{c}\sigma_{d}$, respectively. 
An example is shown in Fig.~\ref{fig:Fig1}(c).
The considered OP therein consists of all two-body terms $J_{ij} s_i s_j$ between four spin variables $s_i$ (black nodes) and is mapped onto six parity spins $\sigma_e$ (colored nodes) together with three parity constraints represented by the two gray triangles and one square.

Second, to map a general parity encoded optimization problem to Rydberg quantum annealers [step two of Eq.~\eqref{eq:overview_transformations}], the three- and four-body constraints are first decomposed into two- and three-body constraints, as illustrated in Fig.~\ref{fig:parity_layouts}(a).
These lower-order constraints serve as the basic building blocks for the Rydberg encoding, and are implemented via the \textsf{LINK} and the \textsf{3BODY} gadgets introduced in Ref.~\cite{Lanthaler2023Ryblo}.
Loosely speaking, the \textsf{LINK} provides the wiring between variables while the \textsf{3BODY} gadget implements the three-body parity constraints.
By assigning appropriate local detunings $\Delta_i$ on the ports of these gadgets, one can encode the OP coefficients $J_{e}$. 

Now we turn to the third step of Eq.~\eqref{eq:overview_transformations}: replacing site-dependent detunings with a uniform global detuning $\Delta$ using the concepts developed in Sec.~\ref{sec:weak_interactions}.
As discussed therein, we are restricted to adjusting detunings strictly on open ports and \textsf{LINK}s; thus, we must first redesign the required detuning patterns.
We do this by developing homogenized gadgets, which are equivalent to the originally proposed gadgets from Ref.~\cite{Lanthaler2023Ryblo} in the sense that they have the same ground-state manifold, but feature uniform weights $\Delta$ on all internal atoms. 
Therefore, all necessary detuning inhomogeneities are shifted to the ports [cf.\ App.~\ref{app:Homogenization}].
Furthermore, the local fields that define the original OP [$J_e$ in Eq.~\eqref{eq:parity_hamiltonian}] translate to additional detunings on the port atoms where the parity variables $\sigma_e$ are encoded.
This results in an atomic layout where local detunings are only required on port atoms.
Figure~\ref{fig:parity_layouts}(b) shows several such layouts, including basic gadgets, and layouts resulting from amalgamation, i.e., the combination of multiple base gadgets.
In these illustrations, we denote the required non-unit detunings without the problem coefficients $J_e$; hereafter, these detunings are denoted as $\Delta_{{\rm port}, i}$ for a given port $i$.
As noted above we have to distinguish two cases of ports:
\textit{(a)} open ports, where we assign a variable of the OP to the state of the corresponding port atom, and \textit{(b)} ports connected via \textsf{LINK}s, where we encode it into the logical states of those, i.e.,
\begin{equation}
    \sigma_ e \mapsto \begin{cases}
\hat{n}_{e} & \text{for open ports},\\
\hat{n}_{\textsf{L}, e} & \text{for \textsf{LINK}s.}
\end{cases}
\end{equation}
where the corresponding operator $\hat{\sigma}_e$ is related to $\hat{n}_{(\textsf{L},) e}$ through
$\hat{n}_{(\textsf{L},) e} = (1 + \hat{\sigma}_e)/2$. Both cases---provided the \textsf{LINK} is long enough---offer enough space for placing the required port or link anchors to realize $\Delta_{{\rm port},e}$ and encode the OP coefficients $J_e$.
This makes the parity-based Rydberg embedding particularly suitable for quantum optimization with globally driven neutral atom arrays.

In the following, we describe in more detail how to determine the anchor positions in both cases as a function of $J_e$, using the ideas presented in Sec.~\ref{sec:weak_interactions}.

\paragraph*{(a) Open ports:}
The first case allows for a straightforward implementation of the required detuning via a nearby anchor atom, as ports situated on the perimeter of the layout have ample surrounding physical space. 
Consider, for instance, the open ports depicted in Fig.~\ref{fig:parity_layouts}(b). 
For such an open port atom $e$, the target detuning comprises two distinct components, the baseline gadget detuning $\Delta_{{\rm port}, e}<\Delta$ and a contribution proportional to the respective optimization problem coefficient $J_e$ [cf.~Eq.~\eqref{eq:spin_glass_OP}]. The total local detuning required at port atom $e$ is thus
\begin{equation}
\Delta_e = \Delta_{{\rm port},e} - \pef J_e \Delta \ ,
\label{eq:total_detuning}
\end{equation}
where $\pef > 0$ represents a dimensionless global normalization constant.
Several remarks regarding Eq.~\eqref{eq:total_detuning} are in order:
First, the negative sign preceding the problem field $J_e$ originates from the relation between the spin variables $\sigma_e$ and the Rydberg occupation numbers $\hat{n}_{(\textsf{L},) e}$. 
Furthermore, the factor of two in this relation is absorbed into the definition of $\pef$. 
Finally, while there is flexibility in choosing $\pef$, its magnitude is restricted by the energy scale at which non-logical states---i.e., states that do not correspond to maximum weight independent sets---become favorable in energy [cf.~discussion in Sec.~\ref{sec:4body}].

To calculate the appropriate anchor position, we directly equate the target local detuning with the effective shift induced by the weak interaction [see Eq.~\eqref{eq:weak_programming}]:
\begin{equation}
\label{eq:weak_eqal_anchor}
\Delta_e(J_e) \overset{!}{=} \Delta_{e}^{\rm weak}(d_{\text{anchor}, e}).
\end{equation}
From this, the required anchor distance $d_{\text{anchor}, e}$ can be determined.
Note that Eq.~\eqref{eq:weak_eqal_anchor} just takes the port atom into account, while ignoring the long-range interactions between the anchor and its neighbours. 
As already mentioned in Sec.~\ref{sec:weak_interactions}, these perturbations can be taken into account leading to slight corrections of the anchor placement.

\paragraph*{(b) \textsf{LINK}s:}

As in the case of an open port, \textsf{LINKs} can encode a logical variable and, thus, demand a logical link detuning corresponding to the respective problem field $J_e$. 
Additionally, as a consequence of gadget amalgamation, the two ports may require local detunings, $\Delta_{\rm port, 1}$ and $\Delta_{\rm port, 2}$, that deviate from the uniform value $\Delta$ provided by the global drive. 
To account for these deviations, we define the port-specific detuning \decifit{}s as:
\begin{equation}
\delta_{\rm port, 1 (2)} := \Delta_{\rm port, 1 (2)}^{}-\Delta .
\end{equation}
Since these \decifit{}s are required on non-open ports near adjacent gadgets, they cannot be induced by placing anchors close to the ports, as doing so would induce detrimental energy shifts on other atoms.
Nevertheless, they can be consolidated into a single local detuning at the center of the \textsf{LINK} due to the non-locality of the logical states $\ket{0}_\textsf{L}$ and $\ket{1}_\textsf{L}$.
In particular, the following detuning-patterns result in the same energy-difference between the logical states (black nodes have detuning $\Delta$):
\begin{align}
\begin{split}
\tikz{
\draw[line width=0.25mm] (0,0) -- (2,0);
\node[ fill, dblu, inner sep=2pt] at (0,0) {};
\draw[fill, dblu] (0,0) node[above, yshift=.03cm, font=\footnotesize, rotate=0]{$\Delta+\delta_{\mathrm{port}, 1}$}; 
\draw[fill] (0.5,0) circle (.07 cm) node[above, yshift=.03cm, font=\scriptsize]{}; 
\draw[fill] (1,0) circle (.07 cm) node[above, yshift=.03cm, font=\scriptsize]{}; 
\draw[fill] (1.5,0) circle (.07 cm) node[above, yshift=.03cm, font=\scriptsize]{}; 
\node[ fill, dblu, inner sep=2pt] at (2,0) {};
\draw[fill, dblu] (2,0) node[above, yshift=.03cm, font=\footnotesize, rotate=0]{$\Delta+\delta_{\mathrm{port},2}$}; 
%
}  & \simeq \hspace{-0.025cm} \tikz{
\draw[line width=0.25mm] (0,0) -- (2,0);
\node[ fill, inner sep=2pt] at (0,0) {}; 
\draw[fill] (0.5,0) circle (.07 cm) node[above, yshift=.03cm, font=\scriptsize]{}; 
\draw[fill, dblu] (1,0) circle (.07 cm) node[above, yshift=.03cm, font=\footnotesize]{$\Delta+\delta_{\mathrm{port},1}+\delta_{\mathrm{port},2}$}; 
\draw[fill] (1.5,0) circle (.07 cm) node[above, yshift=.03cm, font=\scriptsize]{}; 
\node[ fill, inner sep=2pt] at (2,0) {};
}
\\
& \simeq \quad \tikz{
\draw[line width=0.25mm] (0,0) -- (2,0);
\node[ fill, inner sep=2pt] at (0,0) {}; 
\draw[fill] (0.5,0) circle (.07 cm) node[above, yshift=.03cm, font=\footnotesize]{}; 
\draw[fill] (1,0) circle (.07 cm) node[above, yshift=.03cm, font=\footnotesize]{}; 
\draw[fill, dblu] (1.5,0) circle (.07 cm) node[above, yshift=.03cm, font=\footnotesize]{$\Delta-\delta_{\mathrm{port},1}-\delta_{\mathrm{port},2}$}; 
\node[ fill, inner sep=2pt] at (2,0) {};
} \,.
\label{eq:shift_weights}
\end{split}
\end{align}
In addition to the detuning \decifit{}s from the ports, there might be a problem-specific field contribution $- c J_e\Delta$ as in the case of an open port [see Eq.~\eqref{eq:total_detuning}].
This results in a detuning
\begin{equation}
    \Delta_e^{\textsf{LINK}, {\rm anchor}} = \delta_{\rm port , 1}+ \delta_{\rm port , 2} - c J_e\Delta 
\label{eq:total_detuning_link}
\end{equation}
that should be induced by the link-anchor.
Depending on the sign of $\Delta_e^{\textsf{LINK}, {\rm anchor}}$ this can be achieved by a centered or offset anchor as discussed in Sec.~\ref{sec:weak_interactions} [cf.\ Fig.~\ref{fig:weak_interaktion}(c)], by requiring that
\begin{equation}
\label{eq:weak_eqal_link}
    \Delta_e^{\textsf{LINK}, {\rm anchor}} \left(J_e\right) \overset{!}{=} \dwl{e}(d_{{\rm anchor},e}).
\end{equation}
This condition implicitly defines the corresponding anchor distance $d_{{\rm anchor},e}$ from the port detuning \decifit{}s $\delta_{\rm port, 1/2}$ and the optimization problem constant $J_e$.

In Fig.~\ref{fig:parity_layouts}(b, center), an example of this ``shifting'' of detunings according to Eq.~\eqref{eq:shift_weights} is indicated by a dashed arrow, where the gadget amalgamation would require a detuning of $\frac{5}{4}\Delta$ on one of the \textsf{LINK} ports, which is ``shifted'' towards the center of the link and handled by an offset link anchor.

In conclusion, using the homogenized gadgets and the anchor placement strategy introduced above allows us to map arbitrary OPs onto Rydberg atom arrays with a uniform, global detuning $\Delta$.
Furthermore, the homogenized gadgets allow us to apply the inter-gadget crosstalk compensation strategy from Ref.~\cite{Lanthaler2023Ryblo} to account for long-range vdW interactions.
As these required corrections are small and can be fully mapped onto the ports, they are implicitly included in $\Delta_{{\rm port},e}$ throughout this manuscript. 
A detailed explanation of this compensation mechanism is provided in App.~\ref{app:LocalCompensationDetails}.

\section{Experiments}\label{sec:exps}

We now present experimental results for our newly developed approach for solving OPs on globally driven quantum annealers. In particular, we analyze and benchmark the basic building blocks that frequently appear in practical layouts---\textsf{LINK}, \textsf{3BODY}, \textsf{4BODY}, and \textsf{KITE} gadgets [cf.~Fig.~\ref{fig:parity_layouts}(b)]---on a Pasqal Orion Alpha/Fresnel machine~\cite{darcangelo2024, Leclerc2025}. 
Finally, we scale up these components to perform proof-of-concept experimental quantum optimization with globally driven neutral atom arrays for a parity layout consisting of 43 atoms including seven precisely placed anchors.

In the previous sections, we demonstrated how to map OPs onto globally driven Rydberg quantum annealers. That mapping, however, only considered the classical target Hamiltonian $H_{\rm target} = H_{\rm diag}(t=T)$ with $\Delta(T) > 0$, whose ground state(s) encode the solution(s) of the original OP. 
Solving for these ground state(s) requires dynamically driving the Rydberg platform from easy-to-initialize states toward those solutions.
Here, we employ standard Rydberg quantum annealing to find the solutions of $H_{\rm target}$. 
In particular, we initialize all atoms in their ground state $\ket{0}^{\otimes N}$ [see Fig.~\ref{fig:Fig1}(c)] and set the initial global detuning $\Delta(t=0) \equiv \Delta_0 < 0$ at an initially vanishing Rabi frequency $\Omega(t=0) = 0$. 
Then we slowly drive $H_{\rm Ryd}(t)$ using the three-step, piecewise-linear annealing protocol shown in Fig.~\ref{fig:Fig1}(d).
In the protocol, the detuning $\Delta(t)$ evolves as follows: it is first kept at a constant initial detuning $\Delta_0<0$; it is then slowly increased to reach the target detuning $\Delta>0$; and is kept fixed at this value in the final step. 
Meanwhile, the Rabi frequency $\Omega(t)$ is controlled in a different manner: to drive quantum fluctuations, it is first increased to a maximal value $\Omega_{\rm max}$; it is then kept constant; and finally, it is slowly turned off again, such that the final Hamiltonian corresponds to the target one, $H_{\rm Ryd}(t=T) = H_{\rm target}$.
Provided $T$ is sufficiently long and experimental noise is low enough, adiabaticity guarantees that the system reaches a ground state of $H_{\rm target}$, allowing us to extract the OP solution from the measured Rydberg excitation pattern $n_i$ [cf.\ Fig.~\ref{fig:Fig1}(e)].
In practice, this procedure is repeated for a total of $n_{\rm shots}$ shots to accumulate sufficient statistics.

Our choice of a standard annealing protocol is mainly motivated by its simplicity---being defined by only a few meta-parameters [see Fig.~\ref{fig:Fig1}(c) and Tab.~\ref{Tab:Schedule}]---and physical interpretability. 
While such a simple protocol might not yield the theoretical optimum, it allows for the efficient setting of meta-parameters using, for example, numerical simulations. 
In contrast, in-loop optimization of more complex annealing schedules or variational approaches to Rydberg quantum optimization~\cite{Ebadi2022} are significantly more costly, requiring large numbers of experimental runs for the optimization procedure. 
Therefore, this simple choice is sufficient for our goal of validating and benchmarking our new approach on multiple experiments of individual and combined building blocks.

To reduce detrimental effects from long-range vdW interactions and atom position fluctuations, we utilize vdW compensation [cf.~App.~\ref{app:LocalCompensationDetails}] and mitigate radial distance fluctuations between atoms [cf.\ App.~\ref{app:zcomp}] in the creation of the atomic layouts and the anchor placement for all shown experiments, unless otherwise stated.
All relevant experimental parameters are listed in App.~\ref{app:experimental_params} and Tab.~\ref{Tab:Schedule}.

\subsection{\textsf{LINK} gadget}
\label{sec:link}

In the first experiment, we demonstrate and benchmark our core ideas on a stand-alone \textsf{LINK} gadget, as described in Sec.~\ref{sec:weak_interactions}.
In particular, we choose a link of five atoms with two port anchors at a fixed distance $r_{\Delta/2}$ and an additional link anchor to induce a logical field $\dwl{}$ non-locally along the \textsf{LINK}.
The link anchor is placed above the center (left to center) atom for $\dwl{}<0$ ($\dwl{}>0$) and its distance from the link $\danch{}$ is chosen according to the strength of $\dwl{}$ [cf.~Fig.~\ref{fig:weak_interaktion}(c,~d) and Eq.~\eqref{eq:h_link_eff}];
for $\dwl{} = 0$ the link anchor is not required and we do not place it.
We then perform quantum annealing experiments, as described above, for different values of $\dwl{}$ with a fixed annealing schedule.
That means, in particular, that we only change the position of the link anchor throughout the different experiments.

\begin{figure}[th]
\begin{centering}
\includegraphics[width = \columnwidth]{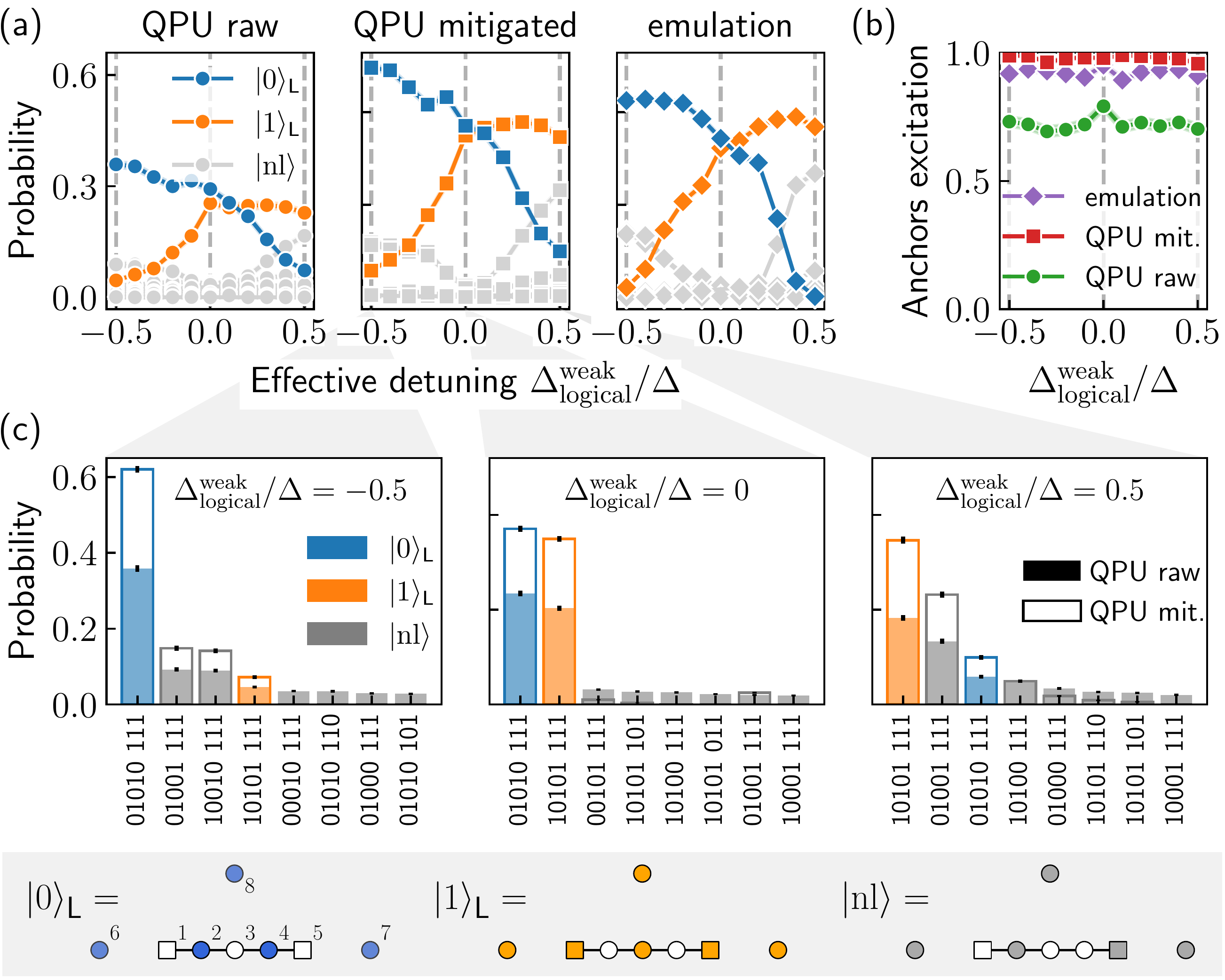}
\par\end{centering}
\protect\caption{\textit{\textsf{LINK} experiments.}
State preparation on a \textsf{LINK} gadget of five atoms plus three anchors for varying logical field $\dwl{}$ induced by the link anchor.
(a)
The probability for measuring the logical states $\ket{0}_{\textsf{L}}$ (blue), $\ket{1}_{\textsf{L}}$ (orange) versus $\dwl{}$ for raw experimental results (``QPU raw''), readout-error-mitigated experimental data (``QPU mitigated''), and noiseless emulation (``emulation'').
Non-logical states are shown as gray lines.
(b)
Anchor excitation probability as a function of $\dwl{}$.
(c)
Bitstring histograms for $\dwl{}/\Delta \in \{-0.5 ,  0.0 , 0.5\}$ (left/center/right). 
Logical states are colored, non-logical states are shown in gray. 
Raw (readout-error-mitigated) experimental data is shown with filled (empty) bars.
Bitstrings are ordered as $\ket{\textsf{LINK}} \ket{\rm{anchors}}$ with the full ordering denoted by the indices in the legend.
In this and all subsequent figures, error bars and shaded bands indicate the standard error.
}
\label{fig:link_exp}
\end{figure}

Our experimental results are summarized in Fig.~\ref{fig:link_exp}.
Figure~\ref{fig:link_exp}(a) shows the probability of measuring the \textsf{LINK} logical states, $\ket{0}_\textsf{L}$ and $\ket{1}_\textsf{L}$ (including anchors in the Rydberg state), as a function of $\dwl{}$.
From left to right, the panels display raw measurement data, readout error-mitigated data [see App.~\ref{app:readout_mitigation}], and noiseless numerical emulations generated via Pasqal's Pulser backend~\cite{LAMBERT20261}.
The raw data exhibit a clear transition: the most probable state changes from $\ket{0}_\textsf{L}$ for $\dwl{} < 0$ to $\ket{1}_\textsf{L}$ for $\dwl{} > 0$, crossing near $\dwl{} \approx 0$.
This aligns qualitatively well with both the noiseless emulations and our theoretical predictions [cf.\ Fig.~\ref{fig:weak_interaktion}(c)].

Despite this qualitative agreement, the raw experimental data yields overall reduced probabilities for measuring the logical states.
This reduction stems from inherent noise of the quantum annealer, such as Rydberg decay, position noise, laser noise, and readout noise~\cite{deLeseleuc2018}.
Let us here focus on readout noise which substantially skews the probabilities for measuring individual bitstrings. 
The effect of readout noise can, however, be partially mitigated during post-processing [cf.\ App.~\ref{app:readout_mitigation}].
We use the tool \textit{M3}~\cite{Nation2021} to perform the mitigation and present the resulting data in the center panel of Fig.~\ref{fig:link_exp}(a).
This mitigated data shows a strong renormalization of the logical state probabilities compared to the raw data and restores good quantitative agreement with the emulation.
A significant portion of this renormalization comes from correcting anchor atoms that fail to be excited to the Rydberg state.
As shown in Fig.~\ref{fig:link_exp}(b), the raw experimental data yields anchor excitation probabilities of only $\approx 70\%$, whereas the corresponding emulation exceeds $\approx 90\%$.
This discrepancy, combined with the relative flatness of the curves as a function of $\dwl{}$, implies that the experimental suppression is driven by intrinsic hardware noise (e.g., readout errors and Rydberg decay) rather than a fundamental limitation of our theoretical concept.
Applying readout error mitigation recovers nearly all anchor excitations, even slightly overestimating them relative to the noise-less emulation.

Figure~\ref{fig:link_exp}(c) displays histograms of the experimentally measured bitstrings for three distinct values of $\dwl{}$.
For larger values of $|\dwl{}|$, non-logical states (depicted in gray) become increasingly prominent.
These states typically contain domain-wall errors, characterized by adjacent atoms simultaneously occupying the ground state.
This observation is in excellent agreement with the system's energy spectrum [see Fig.~\ref{fig:weak_interaktion}(d)], where non-logical states drop to lower energies for sufficiently small anchor distances, $\danch{}$.
Consequently, these erroneous configurations become highly susceptible to thermal or diabatic population 
and, eventually, appear more frequently in the experimental readout.
Nevertheless, even at substantial local fields of $|\dwl{}| = \Delta/2$, the target ground state remains the dominantly measured configuration.
This demonstrates that our approach is capable of sampling the target bitstring even for problems with relatively large local fields.

Furthermore, the center panel of Fig.~\ref{fig:link_exp}(c) shows that at $\dwl{} = 0$, the two logical states are not measured perfectly with identical probabilities, despite those being degenerate ground states for this parameter.
Because the \textsf{LINK} consists of an odd number of atoms, the two logical configurations contain different numbers of Rydberg excitations ($\ket{0}_\textsf{L}$: even, $\ket{1}_\textsf{L}$: odd).
This structural asymmetry, coupled with non-trivial dynamics during the adiabatic annealing protocol, generally leads to unequal sampling probabilities~\cite{Konz2019}.
This disparity is a characteristic of Rydberg-gadget-based architectures---which typically feature target states with differing excitation numbers~\cite{Lanthaler2023Ryblo, Stastny2023}---and is not an artifact of the specific weak-interaction approach introduced here.
Crucially, this sampling bias is sensitive to the exact annealing schedule and can be minimized through parameter optimization to ensure target states are sampled with high probability.

In summary, the results presented in this subsection are a first demonstration that tuning the spatial position of the anchor atoms provides a reliable mechanism to induce local port detunings $\Delta^{\rm weak}_e$ and logical \textsf{LINK} detuning $\dwl{}$ and to accurately program the energy splittings between logical states in globally driven quantum annealers.

\subsection{\textsf{3BODY} gadget}
\label{sec:3body}

\begin{figure}[t]
    \centering
    \includegraphics[width=1\linewidth]{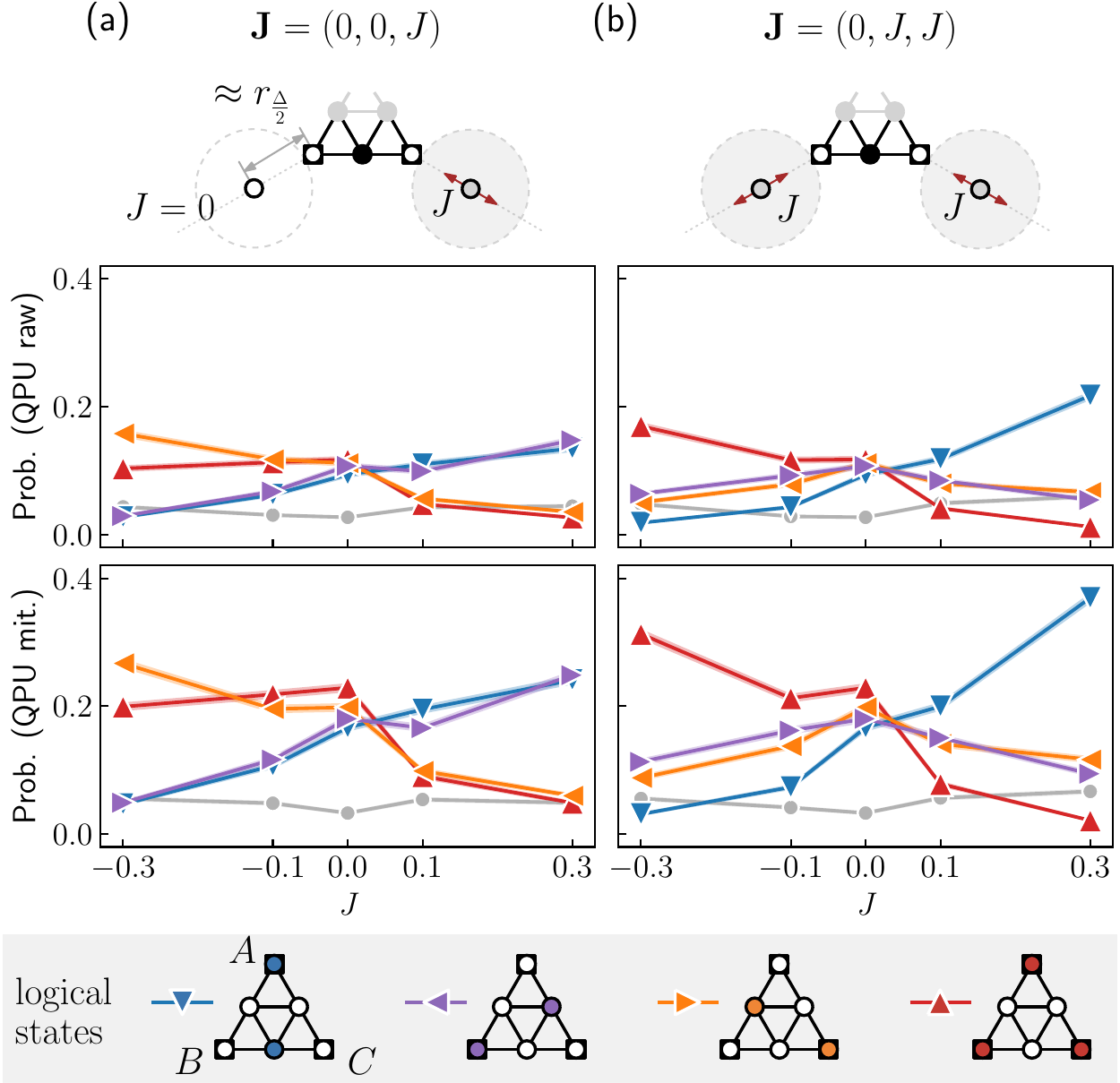}
    \caption{\textit{\textsf{3BODY} experiments.} 
    Probability of measuring logical states after state preparation on a \textsf{3BODY} gadget. Raw experiments (``QPU raw'') and readout-error-mitigated experiments (``QPU mitigated'') are shown from top to bottom.
    The legend shows the four logical states, where colored nodes denote Rydberg excitations and the three auxiliary anchor atoms---all in the Rydberg state---are omitted for clarity.
    (a)
    One problem coefficient $J_C=J$ is varied by changing the position of the corresponding anchor atom, making the red and orange (blue and purple) states energetically degenerate in the final Hamiltonian $H_{\rm target}$, or making all four states degenerate when $J_C=0$.
    (b)
    Two problem coefficients $J_B=J_C=J$ are identically varied by changing the positions of the two corresponding anchor atoms. This leads to unique ground states for $J>0$ (blue state) and $J<0$ (red state) in $H_{\rm target}$.
    In addition to the logical states, we plot in gray the maximum observed population of any non-logical state for each value of $J$.
    }
    \label{fig:Triforce}
\end{figure}

Building upon the successful use of anchor positioning to implement effective detunings in the \textsf{LINK} gadget, we now evaluate this technique within the \textsf{3BODY} gadget, originally proposed in Ref.~\cite{Lanthaler2023Ryblo}.
The \textsf{3BODY} gadget comprises six atoms arranged at the corners and along the edges of a triangle, as depicted in 
in the legend of Fig.~\ref{fig:Triforce}.
The three port atoms $A$, $B$, and $C$ require the port detunings $\Delta_{\mathrm{port}, A/B/C} = \Delta/2$, while the interior atoms are driven with a detuning $\Delta$.
This configuration ensures that the four (quasi-degenerate) lowest-energy states
satisfy the three-body parity constraint ${\sigma_A \sigma_B \sigma_C = 1}$ among the qubits encoded at the ports.
Here, we induce the local port fields by placing three anchor atoms according to Eq.~\eqref{eq:total_detuning}.
We parameterize these fields using the problem coefficient vector $\mathbf{J} = (J_A, J_B, J_C)$.
Modifying $\mathbf{J}$ shifts the position of the respective anchors relative to the baseline distance, $\danch{}(J_e=0) \approx r_{\Delta/2}$ [see Fig.~\ref{fig:parity_layouts}(b) and Fig.~\ref{fig:Triforce}].

Specifically, we conduct two sets of experiments targeting different coefficient vectors to probe distinct physical regimes:
\begin{itemize}
\item[(a)] \textbf{Degenerate:} $\mathbf{J} = (0, 0, J)$, which maintains degenerate ground states as $J$ varies.
\item[(b)] \textbf{Unique:} $\mathbf{J} = (0, J, J)$, which lifts the degeneracy and yields a unique ground state as $J$ varies.
\end{itemize}
The results of these experiments are presented in Fig.~\ref{fig:Triforce}.
Across all runs, we employ a fixed annealing schedule [cf.\ Tab.~\ref{Tab:Schedule}] for all values of $J$ and set the normalization constant $\pef = 1$.
Depending on the specific values of $\mathbf{J}$ different subsets of the four logical states form the ground-state manifold.
For $\mathbf{J} = \mathbf{0}$, all four states are degenerate ground states, since the anchors induce only the baseline detuning $\Delta/2$ equally on all ports. 
The experiment successfully samples all of those dominantly and with non-vanishing, similar probabilities [see Figs.~\ref{fig:Triforce}(a,b), $J=0$].
Experiment (a) introduces a problem field exclusively on the variable $J_C$ associated with port $C$. 
Therefore, the four logical states in the ground-state manifold exhibit two distinct behaviors. 
The logical states with $n_C=1$ (red and orange states in legend of Fig.~\ref{fig:Triforce}), acquire an energy shift of $cJ\Delta$ relative to the other two logical states for which $n_C=0$ (blue and purple).
Therefore, for $J<0$ the red and orange states become the ground states, while for $J>0$, the blue and purple states become the ground states.
In the global detuning approach presented here, $J<0$ ($J>0$) corresponds to placing the anchor atom farther from (closer to) the port, thereby favoring (penalizing) Rydberg excitation for the port atom $C$ through decreased (increased) weak interaction contribution.
The experiments shown in Fig.~\ref{fig:Triforce}(a) successfully sample the appropriate ground states and we clearly see the expected transition from the states with $n_C=1$ (red and orange) being dominantly sampled for $J<0$ to the states with $n_C=0$ (blue and purple) being the most probable ones for $J>0$. 

The blue and purple states are symmetrically equivalent, and appear with equal probability (subject to shot noise).
The red and orange states, on the other hand, feature differing numbers of Rydberg excitations (like the two logical states of the \textsf{LINK}). 
Therefore, although being energetically equivalent in the final Hamiltonian $H_{\rm target}$, they are not expected to be fairly sampled without fine-tuning the annealing sweep~\cite{Konz2019}---an effect clearly visible in our results.

In contrast, experiment (b) applies identical problem fields to ports $B$ and $C$, resulting in three distinct energy levels. The blue state, for which $n_B+n_C=0$, has no energy shift. The purple and orange states, for which $n_B+n_C=1$ acquire an effective energy shift of $cJ$ and remain degenerate. The red state, for which $n_B+n_C=2$, acquires an energy shift of $2cJ$. 
Therefore, for $J<0$, the red state becomes the single ground state, whereas for $J>0$, the blue state becomes the single ground state.
Figure~\ref{fig:Triforce}(b) shows the corresponding experimental state preparation results, where we clearly observe that the respective ground states are the ones being dominantly sampled.
We can also clearly see that the increased gap $\propto |J|$ between ground state and other logical states leads to an improved state preparation performance.

Alongside the raw experimental data, we show readout-error-mitigated results, observing a significant increase in target state probabilities. This indicates that readout errors are a relevant source of experimental noise, but also that their effect can be reduced through classical post-processing.

In summary, these results confirm that our approach successfully extends to the \textsf{3BODY} gadget, and that we can accurately resolve different problem field patterns during state preparation experiments.

\subsection{\textsf{4BODY} gadget}
\label{sec:4body}

\begin{figure}[t]
    \centering
    \includegraphics[width=1\linewidth]{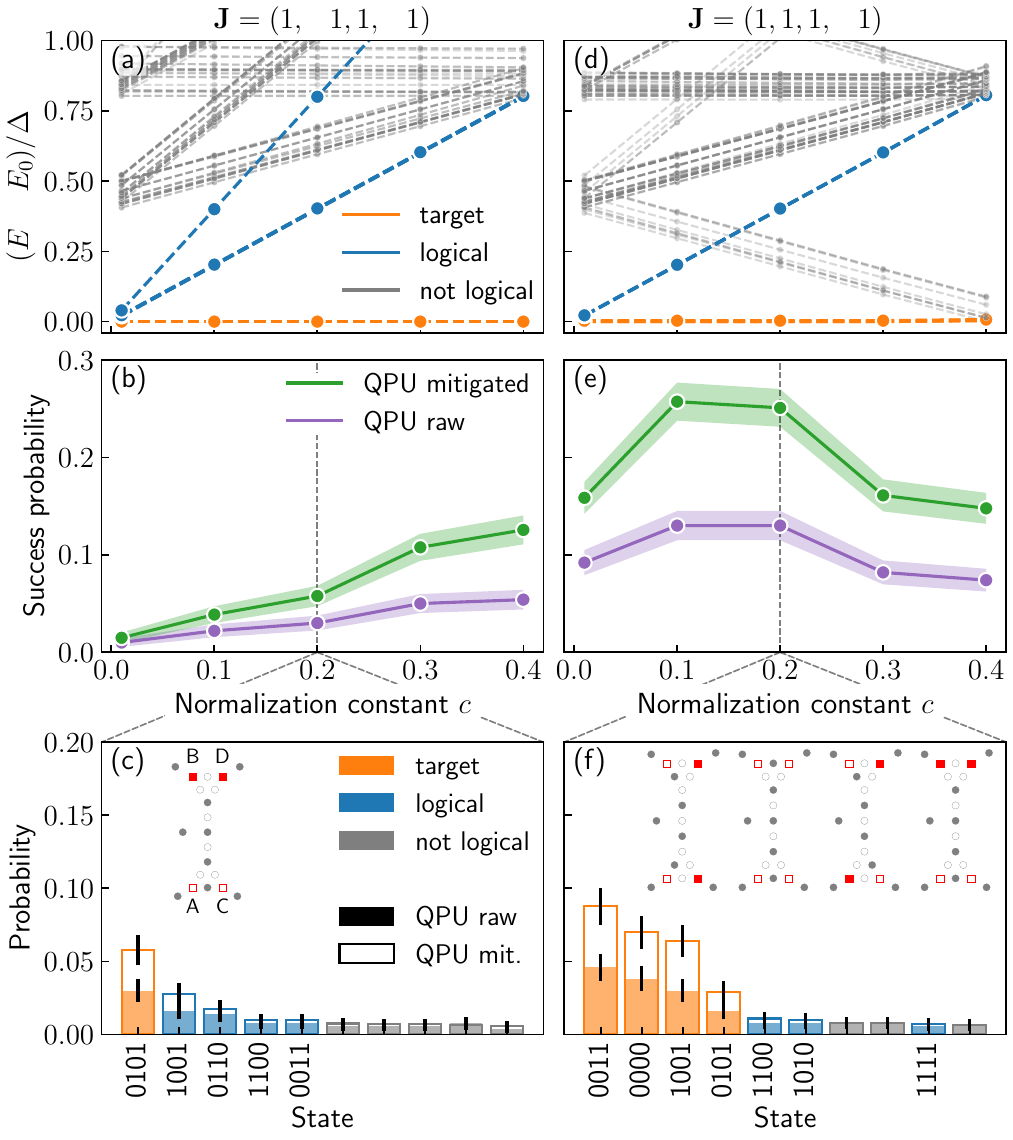}
    \caption{
    \textit{\textsf{4BODY} experiments.}
    We perform state preparation experiments for the \textsf{4BODY} gadget for an unfrustrated (a-c) and frustrated (d-f) set of parameters $\mathbf{J}$, modifying the normalization constant $\pef$. 
    (a, d)
    Low-energy spectra of the final Hamiltonian $H_{\rm target}$ as a function of $\pef$. Each line corresponds to a given bitstring. Target, logical, non-logical bitstrings are depicted in orange, blue, gray respectively.
    We plot the scaled energy gap to the ground state energy $E_0(\pef)$.
    (b, e)
    Total probability of observing the target state(s) as a function of $\pef$. Raw (readout-error mitigated) experimental results are shown in purple (green) color.
    (c, f)
    Bitstring histograms for $\pef=0.2$. Raw (readout-error mitigated) experimental results are shown with filled (empty) bars. We only write partial bitstrings for the port atoms (ordered according to $ABCD$, see inset) for target (orange) and logical (blue) states and omit writing non-logical bitstrings (gray). 
    Filled nodes in the inset represent Rydberg excitations, ports are indicated with red squares.
    }
    \label{fig:4body}
\end{figure}

A standard parity-encoded problem typically involves three- and four-body parity constraints. To achieve planar Rydberg layouts, the four-body constraints are decomposed into two three-body constraints [cf.\ Fig.~\ref{fig:parity_layouts}(a)]. Using Rydberg gadgets, this decomposition is realized by amalgamating two \textsf{3BODY} gadgets via a shared \textsf{LINK} gadget, as illustrated in Fig.~\ref{fig:parity_layouts}(b). Because the weights at the connected ports sum to $\Delta$, this \textsf{4BODY} gadget has uniform detunings of $\Delta$ on all interior atoms, while the four open ports (labeled $A, B, C, D$) obey $\Delta_{\mathrm{port}, A/B/C/D} = \Delta/2$. Additionally, we assign the problem fields $\mathbf{J} = (J_A, J_B, J_C, J_D)$ to the respective open ports.

For our global detuning approach, we introduce the required four anchors at the open ports according to Eq.~\eqref{eq:total_detuning}, alongside an additional link anchor. This link anchor does not encode any problem coefficients; rather, together with compensation terms on the other anchors, it solely compensates for energy shifts arising from the long-range tails of the van der Waals interactions (detailed in App.~\ref{app:LocalCompensationDetails}). Due to its increased size and complexity, the \textsf{4BODY} gadget features a rich energy landscape with qualitatively distinct regimes, making it a versatile benchmark for both our approach and the underlying quantum annealing hardware.

We now consider a series of experiments analyzing the impact of two qualitatively different problem field configurations on the annealing results:
\begin{itemize}
    \item[(i)] \textbf{Unfrustrated}: $\mathbf{J}=(1, -1, 1, -1)$.
    \item[(ii)] \textbf{Frustrated}: $\mathbf{J}=(1, 1, 1, -1)$.
\end{itemize}
We use the following definition to distinguish between frustrated and unfrustrated configurations:
Given a configuration of problem fields $\mathbf{J}$, we define the \textit{field-aligned state} as the one minimizing each problem field locally, without considering any constraints, i.e., the ground state of Eq.~\eqref{eq:parity_hamiltonian} for $\lambda=0$.
On the Rydberg gadget level, this would mean to choose the logical configuration with $\hat{n}_{(\textsf{L},) e} = 0$ ($\hat{n}_{(\textsf{L},) e} = 1$) for $J_e > 0$ ($J_e < 0$) for each open port and \textsf{LINK} individually.
If the \textit{field-aligned state} fulfills all constraints, it remains the ground state for the full problem including the constraint terms, $\lambda >0$.
In this case we term the configuration $\mathbf{J}$ \textit{unfrustrated}.
In contrast, if the \textit{field-aligned state} breaks (some) constraints, it is not the ground state of the full problem with $\lambda>0$.
Instead, the ground state requires an energetic trade-off between fulfilling the constraints and minimizing the local fields.
We term such configurations $\mathbf{J}$ \textit{frustrated}.

Our results for the \textsf{4BODY} gadget are summarized in Fig.~\ref{fig:4body}. The energy spectra for the final Hamiltonian $H_{\rm target}$ are shown as a function of the normalization factor $\pef$ in Figs.~\ref{fig:4body}(a, d) for the unfrustrated and frustrated configurations, respectively. 
Here, the normalization constant $\pef$ acts as a tuning knob, balancing the strength of the problem fields $J_e$ against the strength of the parity constraint, which is implicitly defined by the atomic distance and the Rydberg blockade radius $r_B$ (or, equivalently, $\Delta$ and $C_6$).
The different logical states (colored lines with circles) obey an energy splitting $\propto \pef$ among each other.
This splitting is very narrow for small $\pef$ such that the target states (orange lines) are not well separated from the other logical states (blue lines). 
However, for small $\pef$ non-logical states that break the Rydberg-blockade-induced \textsf{MWIS} pattern (gray lines) remain well separated with an energy gap $\propto \Delta$ in both cases.

Increasing $\pef$ amplifies the importance of the local problem fields, driving a linear increase in the energy splitting between the target states and the other logical states. 
In the unfrustrated case, there is no competition between minimizing the local fields and satisfying the parity constraint; thus, the energy gap between the non-logical states and the target state also grows with $\pef$, making a large $\pef$ ideal. 
In the general frustrated case, however, competition arises. Because the Rydberg gadgets enforce a finite parity constraint strength via the finite blockade strength, constraint-breaking solutions can become energetically favorable if $\pef$ grows too large. 
This is clearly observed in Fig.~\ref{fig:4body}(d), where non-logical states dip lower in energy as $\pef$ increases. 
Specifically, while the gap between the target states and other logical states grows, the gap to constraint-breaking non-logical states shrinks. This dynamic establishes an optimal value $\pef^*$ where the target states are maximally separated from all other states. For the frustrated \textsf{4BODY} gadget presented here, $\pef^* \approx 0.14$ [cf.~Fig.\ref{fig:4body}(d)].

Figures~\ref{fig:4body}(b, e) illustrate the impact of these spectral features on the quantum annealing experiments. Specifically, we plot the success probability—the total probability of measuring the target states—as a function of the normalization constant $\pef$ for a fixed annealing schedule. 
In the unfrustrated case [Fig.~\ref{fig:4body}(b)], the success probability increases monotonically with $\pef$, demonstrating that quantum annealing can efficiently exploit the widened energy gap in the final Hamiltonian when $\pef$ is large. 
Conversely, in the frustrated case, both the raw and readout-error-mitigated data exhibit a peak success probability near $\pef \in [0.1, 0.2]$, aligning well with our spectral analysis. 
Increasing $\pef$ beyond this optimal window causes non-logical states to be sampled more frequently at the expense of the target states.

Interestingly, the overall success probability for the frustrated configuration is generally higher than that of the unfrustrated configuration. This somewhat counterintuitive behavior is largely explained by target state degeneracy: the unfrustrated case possesses only a single target state, whereas the frustrated case features four target states out of the eight logical states. 
The bitstring histograms for $\pef = 0.2$, presented in Figs.~\ref{fig:4body}(c, f), corroborate this by revealing similar individual target state probabilities across both configurations.

In App.~\ref{app:link_length_tuning} we conduct another series of experiments on the \textsf{4BODY} gadget analyzing the impact of changing its link length.

\subsection{\textsf{KITE} gadget}
\label{sec:kite}

Another frequently appearing building block is formed by amalgamating two \textsf{3BODY} gadgets along their baselines. 
The resulting nine-atom gadget [cf.\ Fig.~\ref{fig:parity_layouts}(b)] requires interior local detunings ranging from $2\Delta$ to $4\Delta$, making it incompatible with our global driving approach. 
However, as outlined in App.~\ref{app:Homogenization}, this limitation can be overcome by employing a homogenized version of the gadget, which we refer to as the \textsf{KITE} gadget. 
While retaining the original atomic geometry, the \textsf{KITE} gadget only requires a uniform detuning of $\Delta$ across all interior nodes, with local detunings $\Delta_{\mathrm{port}, e} < \Delta$ restricted to its four ports [see Fig.~\ref{fig:parity_layouts}(b)]. 
Consequently, this homogenized version is directly compatible with our global driving approach.
To implement the gadget as a standalone object under a uniform global drive, we position four programming anchors adjacent to its ports, as depicted in Fig.~\ref{fig:parity_layouts}(b). 
In contrast to the \textsf{3BODY} gadget, not all ports require identical local detunings, making the anchor distances for the left/right ports different from the ones of the top/bottom ports, even if no problem fields $J_e$ are applied.
Another distinct characteristic of the \textsf{KITE} gadget is that its spatial geometry provides only weak isolation between the left/right ports and their neighboring interior atoms.
Consequently, and in contrast to the other gadgets, the detrimental energy shifts induced by the left/right anchors on these neighbors cannot be treated perturbatively.
To mitigate this issue, we incorporate longer-range interactions explicitly into the anchor-placement routine.
Specifically, achieving ideal placement means the different anchors can no longer be treated independently.
Instead, we compute their combined effect on all four logical states of the \textsf{KITE} gadget and optimize their positions simultaneously to match the target behavior.

\begin{figure}[t]
    \centering
    \includegraphics[width=1\linewidth]{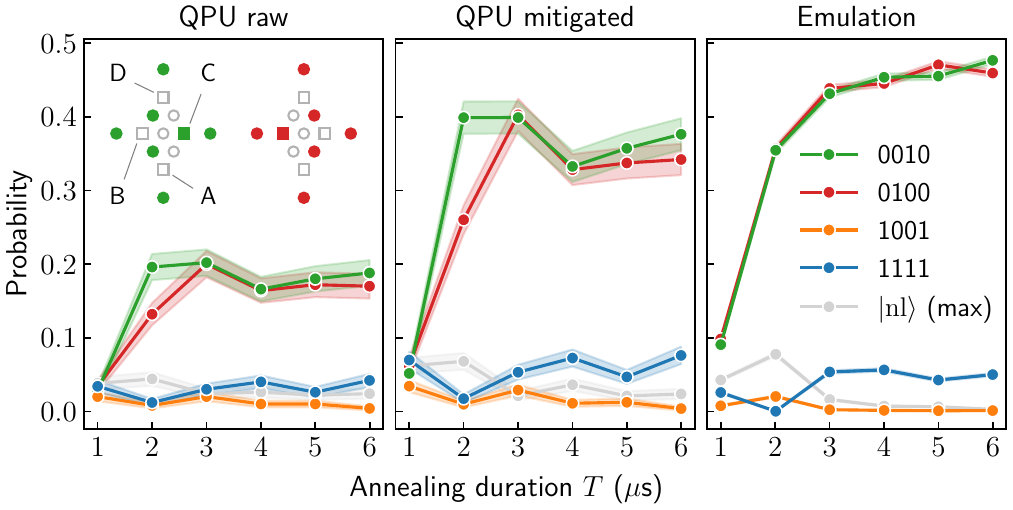}
    \caption{\textit{\textsf{KITE} experiments}.
    We evaluate the performance trade-off between adiabatic state preparation and decoherence across a range of annealing durations $T$ for the \textsf{KITE} gadget with problem coefficients $\mathbf{J}=\left(1, -1, -1, 1\right)$. 
    This specific problem instance features a two-fold degenerate ground-state manifold spanned by the two logical states illustrated in the inset, where colored nodes denote Rydberg excitations. 
    We plot the probability of observing logical states (colored lines) as a function of $T$ for raw and readout-error mitigated experiments, and noiseless emulation from left to right.
    In addition to the logical states, we plot the maximum observed population of any non-logical state for each $T$ (gray line).
    The legend gives partial bitstrings of the port atoms $A, B, C, D$ (see inset) for the logical states.
    }
    \label{fig:kite_experiment}
\end{figure}

We now report on a series of experiments conducted on a single \textsf{KITE} gadget to evaluate the system performance as a function of the annealing duration $T$. 
Figure~\ref{fig:kite_experiment} displays the experimental results for $T$ ranging from $1\,\mu\text{s}$ to $6\,\mu\text{s}$.
In particular, to also show the programmability of the gadget, we apply problem fields $\mathbf{J} = (1, -1, -1, 1)$ that break the ground-state degeneracy and isolate two of the four logical states as target states [cf.~inset of Fig.~\ref{fig:kite_experiment}].
As $T$ increases and the state preparation becomes more adiabatic we observe a sharp initial increase in the experimentally measured population of the two target states [see Fig.~\ref{fig:kite_experiment}, left panel]. 
While noiseless emulations [see Fig.~\ref{fig:kite_experiment}, right panel] suggest that this success probability should monotonically improve with $T$, the experimental data exhibits a stagnation profile instead and plateaus for $T \gtrsim 2 \, \mu{\rm s}$. 
The same trend is also observed for the readout error mitigated results [see Fig.~\ref{fig:kite_experiment}, center panel].
This highlights the competition between improved adiabaticity for larger $T$ and increasing effect of hardware noise sources, such as finite lifetime of the Rydberg state, laser phase and intensity fluctuations, atom position noise, etc.
Nevertheless, up to the maximum available annealing time $T=6\, \mu{\rm s}$ of the hardware, the success probability remains stable, indicating that hardware noise does not overcome the improvements gained from better adiabaticity for such durations.

\subsection{Fully connected graph with four logical variables}
\label{sec:LHZ}

\begin{figure*}[t]
    \centering
    \includegraphics[width=1\linewidth]{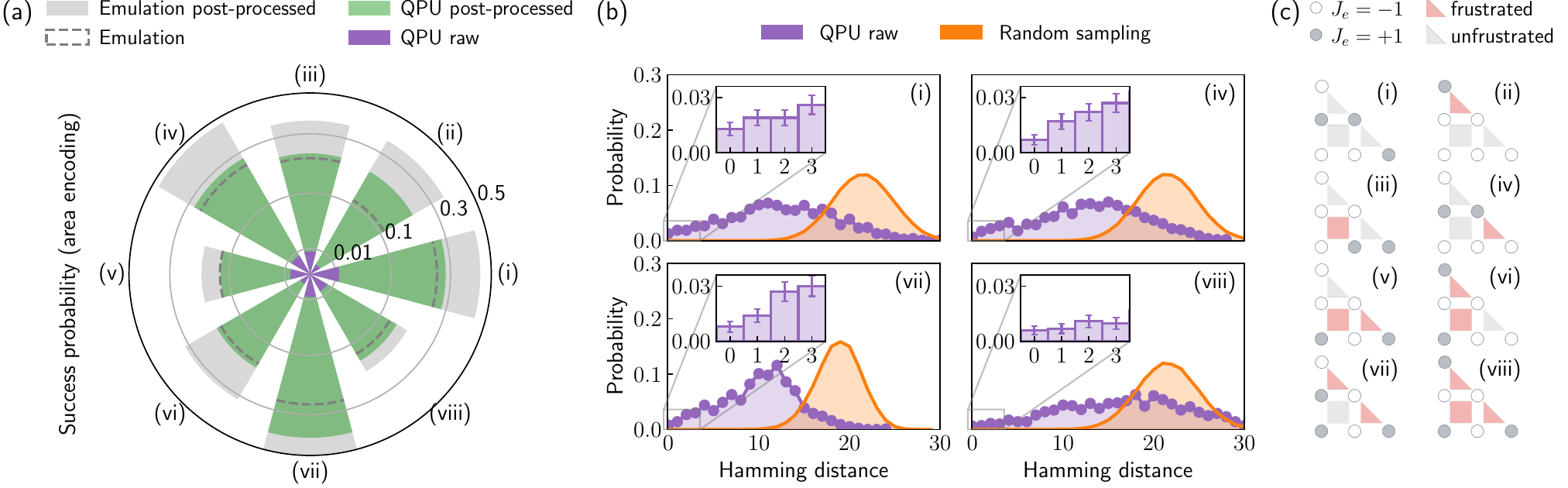}
    \caption{
    \textit{Fully connected QUBO problems}.
    Results from state preparation experiments on the atomic layout shown in Fig.~\ref{fig:Fig1}(c) consisting of 43 atoms including seven anchors.
    This layout corresponds to the parity layout (LHZ triangle~\cite{Lechner2015}) depicted in Fig.~\ref{fig:Fig1}(c) which encodes QUBO problems on a weighted all-to-all graph with four vertices. Different roman numbers correspond to different OP coefficients $\mathbf{J}$ as depicted in (c).
    (a)
    Total probability for measuring the target state(s) for different $\mathbf{J}$, depicted as the area of the individual sectors.
    We compare raw (purple) and post-processed (green) experimental results with noise-less emulation (empty dashed) and post-processed noise-less emulation (gray).
    (b)
    Hamming distance histograms of the experimentally measured bitstring distribution (purple) for four specific choices of OP coefficients, with zero (i), one (iv), two (vii) and three (viii) frustrated parity constraints, whereas all besides (vii) have a single target state.
    The hamming distance is defined as the minimum number of bitflips between a measured bitstring and the target states.
    As reference, we show the distribution of randomly sampled bitstrings in orange.
    (c)
    Different OP coefficients $\mathbf{J}$ depicted in the parity layout. $J_e = -1\ (1)$ is depicted as empty (filled) circles.
    We indicate constraints that are satisfied (frustrated) by the corresponding \textit{field-aligned state} with gray (red) squares and triangles.
    }
    \label{fig:lhz_experiment}
\end{figure*}

Having benchmarked all individual building blocks, in the final part of this work we now investigate a larger layout that integrates all of them.
This 43-atom layout, which includes seven anchors, is depicted in Fig.~\ref{fig:Fig1}(c).
It corresponds to a fully connected QUBO problem on four variables $s_i$, mapping to a parity problem on six parity variables $\sigma_e$ on a triangular LHZ layout~\cite{Lechner2015}.
Six of the anchors directly induce the problem fields $J_e$ as effective local detunings on the respective open ports or \textsf{LINKs}, while the seventh accounts for the port detuning \decifit{} arising from the amalgamation of the \textsf{KITE} gadget with the vertical \textsf{LINK} gadget [cf.\ Fig.~\ref{fig:parity_layouts}(b), gray anchor].

For this final series of experiments, we employ long annealing durations of $T=6\, \mu\mathrm{s}$ [cf.\ Sec.~\ref{sec:kite}] to minimize the diabatic errors that are particularly important for larger layouts.
Furthermore, to ensure thorough benchmarking, we implement eight distinct problem instances with problem fields $J_e \in \{+1, -1\}$.
Each instance is representative of an entire class of optimization problems, defined by its degree of frustration in the parity-mapped (LHZ) problem (i.e., which parity constraints are violated for the \textit{field-aligned state} that locally minimizes each individual problem field $J_e$ [cf.\ Sec.~\ref{sec:4body}]).
We distinguish the following eight classes [cf.~Fig.~\ref{fig:lhz_experiment}(c)] which we divide into four categories defined by the number of frustrated constraints in the field-aligned state:
(i) Unfrustrated; 
(ii--iv) one parity constraint violated;
(v--vii) two parity constraints violated;
(viii) Fully frustrated---all three parity constraints violated.

Figure~\ref{fig:lhz_experiment}(a) shows the success probability---the total probability of measuring the target state(s)---for the eight considered instances.
Across all instances, the raw experimental data (purple slices) yields the target state(s) with non-vanishing probability.
Furthermore, with the exception of instances (ii) and (vi), the target states are the most frequently measured bitstrings.
For comparison, we perform noiseless emulation using a matrix-product-state-based emulator available via the Pasqal cloud~\cite{pasqal_cloud_2026}.
The emulation (empty dashed slices) yields significantly higher success probabilities than the raw experimental data, underscoring the severe impact of experimental noise on such larger-scale layouts.
This sensitivity is expected: because we are analyzing exact bitstring probabilities in Fig.~\ref{fig:lhz_experiment}(a), already a single bit-flip on any atom results in a failed measurement.

Given this sensitivity to single bit-flips, we further analyze the sampled states by examining their Hamming distance from the target state(s) in Fig.~\ref{fig:lhz_experiment}(b).
Compared to a baseline of uniformly sampled bitstrings (orange line), the experimentally obtained distribution (purple dots) is heavily skewed toward lower Hamming distances, maintaining substantial weight near the target state(s) [see insets].
This demonstrates that the quantum annealer successfully drives the system toward the target manifold despite the presence of noise, and motivates the use of a dedicated post-processing strategy to mitigate residual bit-flips.
Applying the readout-error mitigation strategy utilized earlier in this work proves ineffective here, owing to the large Hilbert space, low measurement counts, and the resulting flat bitstring distributions [cf.~App.~\ref{app:readout_mitigation}].
Instead, we apply a newly developed post-processing method tailored specifically for gadget-based Rydberg layouts~\cite{cluster_decoder_in_prep}.
This method efficiently corrects for bit-flip errors within small, finite regions of the layout by exploiting Rydberg blockade information.
The post-processed experimental results, depicted by the green slices in Fig.~\ref{fig:lhz_experiment}(a), demonstrate that this method compensates for a significant portion of the hardware errors.
The post-processed experimental success probabilities reach or even surpass the noiseless emulation results, suggesting that the post-processing partially mitigates also diabatic errors.
Applying this same post-processing to the emulated data (gray slices) further increases the emulated success probabilities, corroborating this hypothesis.

These results demonstrate, on a small QUBO instance, that even non-natively embeddable optimization problems can be solved effectively with globally driven quantum annealers. In particular, using anchors to induce the required local detunings is a viable and scalable pathway for hardware limited to global detuning control.

\section{Summary and Outlook}
\label{sec:conclusion}

In this paper, we have introduced and implemented a method for quantum optimization on globally driven neutral atom arrays by utilizing weak Rydberg interactions. 
By leveraging the precise position control of optical tweezers in neutral atom platforms, we precisely place additional anchor atoms to induce effective detunings on individual atoms through energy modulations from weak Rydberg interactions.
We applied this approach to Parity-encoding-based Rydberg embeddings of optimization problems~\cite{Lanthaler2023Ryblo}, showing that its structure is ideally suited for this weak-interaction approach.
Within this framework, the required local detunings can be seamlessly shifted to open ports or extended \textsf{LINK} gadgets, ensuring ample space for anchor placement.

These results eliminate the experimental overhead of local addressability for quantum optimization on neutral atom platforms, significantly simplifying the scaling to larger devices. 
We validated our approach on a Pasqal Orion Alpha/Fresnel Rydberg quantum annealer~\cite{darcangelo2024, Leclerc2025}, confirming the performance of all fundamental building blocks. 
Finally, we integrated these elements in a proof-of-principle 43-atom Rydberg layout, incorporating seven precisely placed anchor atoms, to successfully solve an all-to-all QUBO problem of four variables using purely global driving.


Several improvements to the state preparation method and the hardware could be foreseen to improve our results further, providing a viable path to scale up problem sizes.
Distance fluctuations arising from finite atomic motion limit the resolution of optimization problem fields that can be implemented via weakly interacting anchors. 
Such fluctuations could be suppressed by cooling the atoms closer to the motional ground state~\cite{Kaufman2012, Kaufman_2021, Bluvstein2025, Brown2019}.
Increasing laser power and stability would reduce laser-induced noise during the state preparation and allow to decrease annealing durations due to increased Rabi frequencies, which in turn reduces noise from Rydberg decay.
On the algorithmic side, the pulse shapes used for the adiabatic sweeps in this paper could be further improved using more complex annealing protocols~\cite{GueryOdelin2019, Perseguers2025} or optimal control based approaches~\cite{KHANEJA2005, Ebadi2022, Finzgar2024, Leclerc2025}, with the goal of enhancing the fidelity or reducing the required time budget.

Depending on the specific strengths and features of the neutral atom quantum annealing hardware, it may be advantageous to combine the here presented global-detuning approach with the local-detuning based approach~\cite{Lanthaler2023Ryblo} such that required local detunings can be either induced via precisely placed anchor atoms, or directly via locally controlled laser detunings.
Future research could identify under which conditions the algorithmic performance of such a combined implementation is best, as well as exploring experimental realizations.

In this work, we considered only 2D atomic layouts although our approach can be naturally extended to three-dimensional arrays. In particular, the \textsf{4BODY} gadget can be alternatively implemented using a 3D tetrahedral structure~\cite{Lanthaler2023Ryblo}, with the here presented ideas eliminating the need for 3D local detuning control. 
Additionally, in 3D layouts the positions of the anchor atoms are not restricted to a planar geometry with the potential for further reducing anchor placement restrictions.

Furthermore, we have only considered the Parity-based embedding for optimization problems into Rydberg quantum annealers in this work, given that its structure aligns particularly well with our weak interaction approach. 
Future work could explore extending the weak-interaction approach developed here to other embeddings, such as the crossing lattice~\cite{Nguyen2023} and the quantum wire~\cite{Kim2022}, as well as to a wider range of functionally complete Rydberg gadgets~\cite{Stastny2023}.

In this manuscript, we have solely focused on weak interactions from additionally placed anchor atoms without altering the basic gadget geometries proposed in Ref.~\cite{Lanthaler2023Ryblo}.
However, slightly modifying the positions of atoms within the gadget instead could also induce weak interactions on nearby atoms and induce effective local detunings, enabling more atom-efficient geometries [see App.~\ref{app:weak_without_anchors} for an example].

Finally, it could be instructive to consider (dynamically) induced weak interactions from precisely placed atoms as an alternative mechanism for the implementation of gates for digital neutral atom based quantum computing with the potential to reduce control requirements.

\begin{acknowledgements}
We thank the company Pasqal for providing access to their neutral atoms quantum computer Pasqal Orion Alpha/Fresnel.
We thank Louis Vignoli for insightful discussions, guidance on the quantum computer usage, and providing snapshots of experimental data.
We thank Berend Klaver and Jaewook Ahn for valuable discussions. 
We also thank JuYoung Park for helpful feedback on the manuscript.
This research was funded in whole, or in part, by the Austrian Science Fund (FWF) SFB BeyondC Project No.\ F7108-N38 (DOI: 10.55776/F71). 
This project was supported by a FFG Funding (Project No.\ FO99918691) as part of the international Eureka cooperation, and by the Austrian Research Promotion Agency (FFG Project No. FO999937388, FFG Basisprogramm). 
This research is funded by the German Federal Ministry of Research, Technology and Space (BMFTR) within the project MUNIQC-ATOMS (Project No.\ 13N16080).
This publication has received funding under Horizon Europe programme HORIZON-CL4-2022-QUANTUM-02-SGA via the project 101113690 (PASQuanS2.1).
P.I.\ has received funding from the European Union (Horizon-MSCA-Doctoral Networks) through the project QLUSTER (HORIZON-MSCA-2021-DN-01-GA101072964).
For the purpose of open access, the author has applied a CC BY public copyright license to any Author Accepted manuscript version arising from this submission.
\end{acknowledgements}

\section*{Data availability}
The data that support the findings of this study are available from the corresponding author upon reasonable request.


%

\clearpage

\appendix

\section{Experimental details}
\label{app:experimental_params}

We have used Pasqal's Orion Alpha/Fresnel device~\cite{darcangelo2024, Leclerc2025} for the presented experiments, which we have accessed through the Pasqal cloud \cite{pasqal_cloud_2026}.
The $^{87}$Rb-machine features the following fixed atomic parameters:
$\ket{0}=\ket{5S_{1/2}, F=2, m_F=2}$, 
$\ket{1}=\ket{60S_{1/2}, m_J=1/2}$,
$C_6=138$ GHz $\mu$m$^6$.
Furthermore, we set the standard atomic distance between non-anchor atoms to $a=5 \, \mu{\rm m}$ for all experiments.
Other parameters are experiment specific and we comprehensively list them in Tab.~\ref{Tab:Schedule}.

In all plots of experimental and emulation data the standard error is indicated with error bars or shaded bands.

\begin{table*}[t]
\centering
\caption{Parameters} 
\begin{ruledtabular}
\begin{tabular}{@{}l|cccccccc@{}}
Figure & Fig.~\ref{fig:weak_interaktion} & Fig.~\ref{fig:link_exp} & Fig.~\ref{fig:Triforce} &  Figs.~\ref{fig:4body},~\ref{fig:sigma_z_compensation} & Fig.~\ref{fig:kite_experiment}& Fig.~\ref{fig:lhz_experiment} & Fig.~\ref{fig:4body_link_length} & Fig.~\ref{fig:link_interatomic_distance_change} \\
Gadget / layout & \textsf{LINK} & \textsf{LINK} & \textsf{3BODY} & \textsf{4BODY} & \textsf{KITE} &\textsf{LHZ} & \textsf{4BODY} & \textsf{LINK} \\
%
%
\hline
%
$\Delta_0 / (2\pi)$ [MHz]& -- & $-3.18$ & $-1.65$ & $-1.65$ & $-3.18$ & $-1.65$ & $-1.65$ & -- \\
%
$\Delta / (2\pi)$ [MHz]& $4.41$ & $4.41$ & $4.41$ & $4.41$ & $4.41$ & $4.41$ & $4.41$ & $2.39$ \\
%
$\Omega_{\rm max} / (2\pi)$ [MHz] & -- & $1.43$ & $1.43$ & $1.43$ & $1.43$& $1.43$ & $1.43$ & -- \\
%
$t_{\rm{rise}}$ [$\mu$s]  & -- & $0.3$ & $0.5$ & $0.45$ & $0.19-1.12$ & $1$ & $0.6$ & -- \\
%
$t_{\rm{steady}}$ [$\mu$s] & -- & $0.9$ & $4$ & $3.1$ & $0.81-4.86$ & $4$ & $4.8$ & -- \\
%
$t_{\rm{fall}}$ [$\mu$s]  & -- & $0.8$ & $0.5$ & $0.45$ & $0.19-1.12$ & $1$ & $0.6$ & -- \\
\hline
$\pef$ & -- & $1$ & $1$ & $0.01 - 0.4$ & $0.1$ & $0.2$ & $0.4$ & -- \\
$n_{\rm shots}$ & -- & $\ge 2000$ & $3500$ & $500$ & $500$ & $1000$ & $1000$ & -- \\
\end{tabular}
\end{ruledtabular}
\label{Tab:Schedule}
\end{table*}

\section{Tuning the \textsf{LINK} length}
\label{app:link_length_tuning}

In the experiments presented in the main text, we use a standard link length of five atoms. 
However, this could in general be increased to further enlarge the separation between the gadgets, thereby reducing their crosstalk~\cite{Lanthaler2023Ryblo} and better isolating link anchors from the connected gadgets.
Furthermore, in general layouts, longer links can arise from the structure of the optimization problem [cf.\ Fig.~\ref{fig:parity_layouts}].
To evaluate the impact of varying the link length, we conduct another series of experiments on the \textsf{4BODY} gadget. 
Specifically, we investigate whether longer links between the two \textsf{3BODY} gadgets can faithfully preserve the intended copy functionality and maintain robust correlations across the layout, particularly given limited annealing durations and inherent noise.

\begin{figure}[t]
    \centering
    \includegraphics[width=1\linewidth]{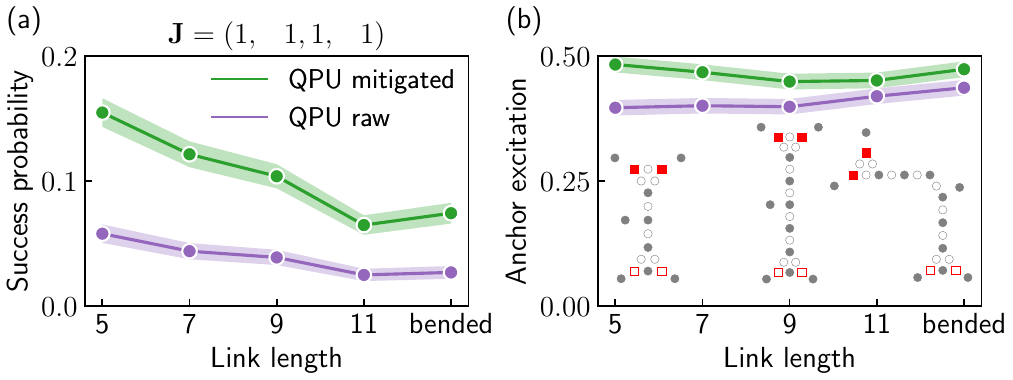}
    \caption{\textit{Tuning the link length in the \textsf{4BODY} gadget.}
    (a) Observed success probability of measuring the target state after adiabatic state preparation of the \textsf{4BODY} gadget across different link lengths. 
    Raw experimental results and readout error mitigated results are displayed in purple and green, respectively. 
    (b) Relative frequency of finding all anchors successfully excited.
    Both panels include comparative data for a geometrically bended link structure, as illustrated in the inset. 
    Within the inset, colored nodes denote the Rydberg excitations corresponding to the target state.
    }
    \label{fig:4body_link_length}
\end{figure}

Figure~\ref{fig:4body_link_length} displays the corresponding results for increasing the link length from the previously used five to eleven atoms.
Figure~\ref{fig:4body_link_length}(a) shows the success probability of preparing the correct ground state for the unfrustrated instance across the different link lengths. 
Again, we plot raw experimental results and readout error mitigated data. 
As the length of the link increases, we observe a gradual decline in the success probability across the raw, mitigated, and post-selected data. 
Successfully preparing the target state on the full layout requires the link to act as a faithful copy gadget, which necessitates spreading information across its entire length. 
In particular, the copying mechanism relies on the creation of the logical states $\ket{0}_\textsf{L}$ and $\ket{1}_\textsf{L}$, which are long-range correlated. 
For a finite annealing duration, however, correlations can only spread a finite distance---the correlation length. 
Experimental noise can further suppress or even arrest this spreading~\cite{Cheneau2012, Marcuzzi2017, Keesling2019, Bombieri2024}. 
Therefore, the observed decrease in success rate can be directly attributed to a finite correlation length induced by the fixed annealing duration, compounded by noise-induced localization phenomena.
However, it is important to note that even for the longest considered links with eleven atoms the target state is the one dominantly sampled, reaching around $6\%$ of all shots when readout errors are mitigated, even without sophisticated optimization of annealing parameters.

Additionally, we investigate a modified \textsf{LINK} geometry, termed ``bended'', which appears frequently in larger layouts as connectors [see inset in Fig.~\ref{fig:4body_link_length}(b)].
Crucially, we observe that this bended variant with eleven \textsf{LINK} atoms exhibits almost identical performance to its straight counterpart, underlining the flexibility of the link in terms of geometric routing.

To complement this analysis, Fig.~\ref{fig:4body_link_length}(b) shows the probability of finding all anchors excited as a function of the link length. 
We observe a nearly constant behavior, indicating that the link length does not strongly affect the excitation of the anchors.

\section{Local compensation strategy}
\label{app:LocalCompensationDetails}

Rydberg gadgets provide a flexible toolkit to encode optimization problems. 
The gadgets themselves form small instances of
\textsf{MWIS} problems, which can be combined to encode more complex instances. 
A Hamiltonian description of these \textsf{MWIS} problems is provided in the main text [Eq.~\eqref{eq:MWISHamiltonian}], where the independent set condition is enforced by the second term, $U \sum_{(i,j)\in E}\hat n_i\hat n_j$, where $U$ is the associated penalty strength $U$.

In the Rydberg platform, this independence constraint is physically implemented via interactions between Rydberg states. 
We encode the logical problem so that an atom in the Rydberg state $\ket{1}$ belongs to the independent set.
Two nearby atoms in the Rydberg state, i.e., both selected for the independent set, experience a large penalty from strong vdW interactions scaling as $V_{\mathrm{vdW}} \propto \vert{}\mathbf{x}_i-\mathbf{x}_j\vert{}^{-6}$, where $\mathbf{x}_i$, $\mathbf{x}_j$ denote the atoms' positions.  
Conversely, atoms separated by large distances can be simultaneously excited without mutual interference, thus both can belong to the independent set.
However, the continuous vdW interaction lacks a sharp cutoff to perfectly mimic the discrete presence or absence of a graph edge in the original \textsf{MWIS} problem. 
To bridge this gap, one defines a threshold distance that separates the interaction into two operational regimes. 
In particular, the continuous potential $V_\mathrm{vdW}(r)$ is approximated as a step potential at the static blockade radius $r_B = (C_6/\Delta)^{1/6}$ (assuming site-independent detunings for simplicity):
\begin{equation}
V_\mathrm{vdW}(r)\approx
\begin{cases}
U & r < r_B, \\
0 & \text{otherwise.}
\end{cases}
\label{eq:BlockadeApproximation}
\end{equation}
Under this step-potential approximation, the Rydberg Hamiltonian in Eq.~\eqref{eq:RydbergClassical} directly reduces to the target \textsf{MWIS} Hamiltonian in Eq.~\eqref{eq:MWISHamiltonian}. 
From a compilation perspective, this mapping allows one to first formulate optimization tasks as \textsf{MWIS} problems and subsequently solve them on a Rydberg platform. 
Nevertheless, this idealization introduces a systematic error, as the long-range tails of $V_{\rm vdW}$ beyond $r_B$ are neglected by the approximation in Eq.~\eqref{eq:BlockadeApproximation}.

Here, we review and refine the local compensation strategy introduced in Ref.~\cite{Lanthaler2023Ryblo}, which addresses this systematic error. 
In this approach the effect of vdW contribution can be counteracted by adapting some of the local detunings 
\begin{equation}
    \Delta_i \mapsto \tilde{\Delta}_i = \Delta_i + \Delta_i^{\rm comp}
\end{equation}
with compensation terms $ \Delta_i^{\rm comp}$.
To describe this compensation procedure in detail, it is helpful to distinguish between two categories of gadgets.

The first category includes copy gadgets, such as the \textsf{LINK}, whose primary purpose is to encode logical states in a non-local manner. 
Copy gadgets are characterized by the property that fixing the configuration on one port uniquely determines the configurations on all remaining ports.

The second category comprises modules, i.e., gadgets that enforce more complex constraints on their port variables, such as the \textsf{3BODY} or \textsf{KITE} gadgets. 
Unlike copy gadgets, fixing the configuration on one port of a module still leaves internal degrees of freedom across the remaining ports. 
These varied excitation patterns can induce crosstalk in the form of detrimental, configuration-dependent energy shifts between adjacent modules. 
To mitigate this issue, Ref.~\cite{Lanthaler2023Ryblo} proposed introducing intermediate \textsf{LINK} gadgets to spatially isolate modules to a distance where inter-module crosstalk becomes negligible. 
Residual interactions within a given module, as well as between modules and their connecting links (copy gadgets), can then be managed by considering the effect of neighbouring gadgets only, as discussed below.
The specific modules and copy gadgets relevant to this work are illustrated in Fig.~\ref{fig:homogenization}. 
Copy gadgets include the \textsf{LINK} and the three-port \textsf{FORK} gadget, which enforces identical configurations across all of its ports. 
Modules include the \textsf{3BODY} and \textsf{KITE} gadgets.

Consider a \textsf{UD-MWIS}-encoded parity optimization problem composed of an amalgamation of modules and copy gadgets.
Furthermore, consider the bare layout without local fields, $J_e = 0, \ \forall e$, where the detunings account solely for the baseline contributions $\Delta_{\mathrm{port}, e}$ required to construct the gadgets. 
In this case, any logical state (i.e., any state corresponding to a valid \textsf{MWIS} solution) solves the problem. 
Consequently, the logical subspace should be degenerate: for any two logical states $\vert{}n\rangle = \vert{}n_1 n_2 \dots\rangle$ and $\vert{}m\rangle = \vert{}m_1 m_2 \dots\rangle$, $n_i,m_i\in\{0,1\}$, the expectation values must satisfy $\langle n \vert{} H_{\mathrm{diag}} \vert{} n \rangle = \langle m \vert{} H_{\mathrm{diag}} \vert{} m \rangle$. 
In other words, the condition
\begin{equation}
\sum_{i,j} V_\mathrm{vdW}(r_{ij})(n_i n_j - m_i m_j) = \delta E_{m,n} \overset{!}{=} 0
\label{eq:VanishingTailsCondition}
\end{equation}
must hold for all logical states. 
Note that there is no contribution proportional to $\sum_i \Delta_i \hat{n}_i$ as both states $\vert{}n\rangle$ and $\vert{}m\rangle$  correspond to maximal sets with equal total weight. 
Furthermore, we define 
$r_{ij} = {\vert{}\mathbf{x}_i-\mathbf{x}_j\vert{}^{-6}}$ as a shorthand notation for the distance between atoms $i$ and $j$.

The condition in Eq.~\eqref{eq:VanishingTailsCondition} is challenging to satisfy in general and typically holds only for specific atomic arrangements. 
In the following, we introduce additional physical degrees of freedom via supplementary detunings $\Delta_i^{\rm comp}$ and show how to (approximately) enforce the degeneracy of logical states by choosing the $\Delta_i^{\rm comp}$ appropriately.

\begin{figure}[t]
    \centering
    \includegraphics[width=1\linewidth]{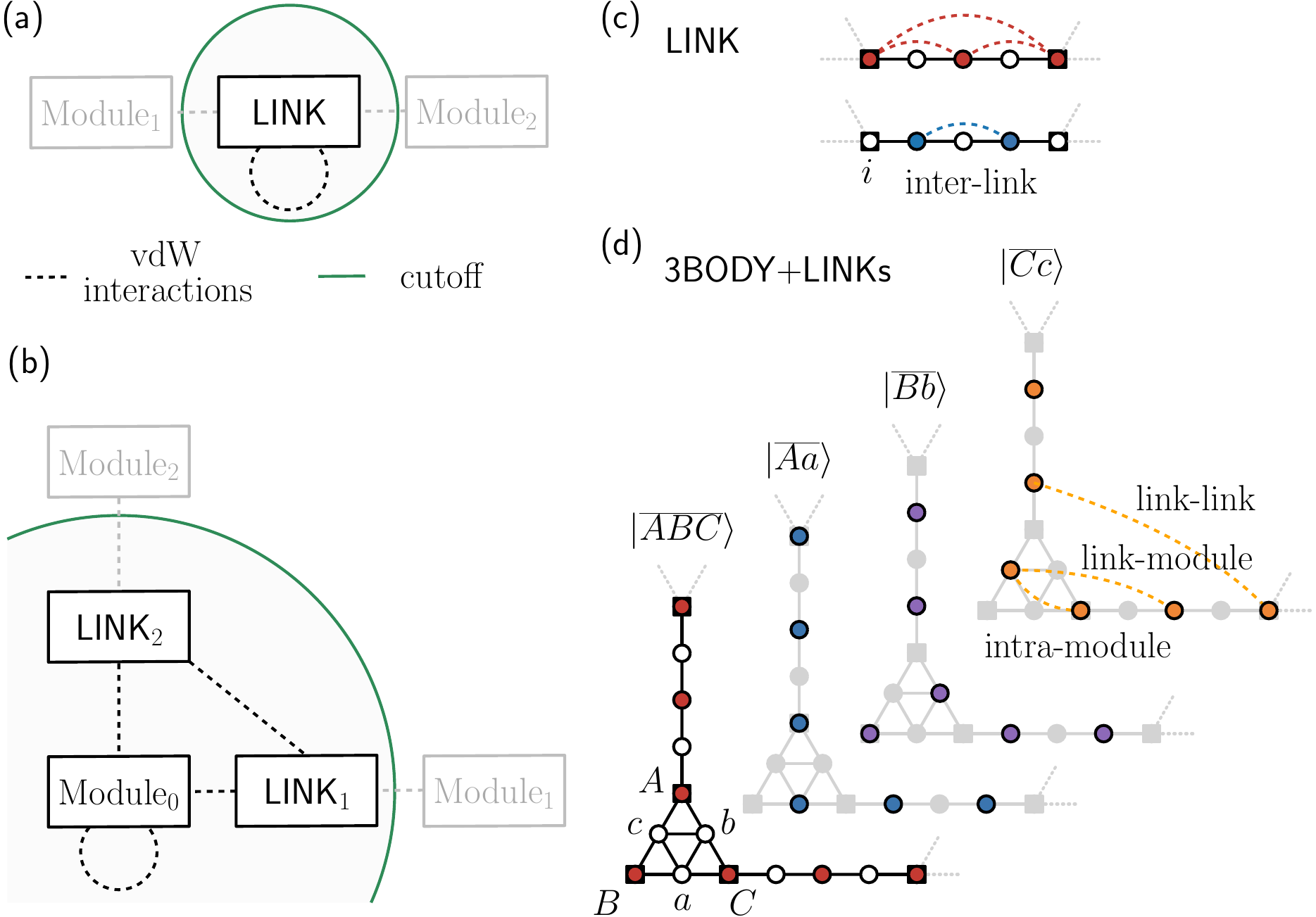}
    \caption{
    \textit{Local compensation:} State-dependent energy shifts resulting from long-range vdW interactions are handled on a neighborhood basis, requiring the inclusion of at most directly adjacent gadgets.
    (a) For a \textsf{LINK} (copy gadget), only intra-gadget contributions are taken into account, which are represented here as self-loops.
    (b) For modules—such as the \textsf{3BODY} gadget—adjacent \textsf{LINK} gadgets and the interactions between them are included.
    (c) In the case of a \textsf{LINK} gadget, the compensation term $\Delta_i^{\rm comp}$ on port $i$ is determined by calculating the vdW energy contribution for the logical states $\vert{}1\rangle_{\textsf{L}}$ (red) and $\vert{}0\rangle_{\textsf{L}}$ (blue).
    (d) Similarly, for modules, intra-module, link-module, and link-link (or general copy gadget) interactions must be summed up to define the compensation detunings on nodes $a, b,$ and $c$, as detailed in Appendix~\ref{app:LocalCompensationDetails}. 
    Different modules require distinct compensation terms, as depicted in Fig.~\ref{fig:homogenization}. 
    As an example, for the \textsf{3BODY} gadget shown here, the compensation term $\Delta_{a}^{\rm comp}$ on atom $a$ is determined by the vdW contributions of state $\vert{}\overline{Aa}\rangle$ relative to the reference state $\vert{}\overline{ABC}\rangle$.
    }
    \label{fig:compensation}
\end{figure}

To this end, we assume the module-separating links are long enough such that module-module interactions in Eq.~\eqref{eq:VanishingTailsCondition} can be safely neglected (see Ref.~\cite{pichler2018computationalcomplexityrydbergblockade} for rigorous bounds on the energy contributions of distant atoms). 
Rather than adhering strictly to a rigid spatial cutoff radius, we practically include all copy gadgets attached to a given gadget while neglecting atoms that belong exclusively to other gadgets as depicted in Fig.~\ref{fig:compensation}(a,b).
This inclusion of neighbouring copy gadgets is motivated by the fact that knowing the logical state within a gadget uniquely determines its continuation along the attached copy gadgets.

In the case of copy gadgets, no adjacent gadgets are included; the remaining terms consist solely of intra-gadget vdW interactions, as depicted in Fig.~\ref{fig:compensation}(c). 
These contributions can be compensated by evaluating them for the two logical states. 
Because the logical state information is distributed globally across the entire copy gadget, this compensation can technically be assigned to any individual atom, provided the proper sign is taken into account.
In Fig.~\ref{fig:compensation}(c), we choose port atom $i$ as an illustrative example, which receives the additional compensation term $\Delta_i^{\rm comp}$. 
Alternatively, one can select multiple atoms or particularly, one or more ports and distribute the required detuning weight among them.

To illustrate the compensation mechanism for modules, let us consider the representative case of a \textsf{3BODY} gadget. 
We denote the three port atoms of the \textsf{3BODY} gadget as $A$, $B$, $C$ and the auxiliary atoms as $a$, $b$, $c$. 
Furthermore, we label the four logical states as $\vert{}ABC\rangle$, $\vert{}Aa\rangle$, $\vert{}Bb\rangle$, and $\vert{}Cc\rangle$. 
Here, the state label indicates all atoms that are Rydberg excitated 
as showcased in Fig.\ref{fig:compensation}(d).
Crucial to the vdW compensation is the observation that these four logical states are uniquely distinguished by their excitation patterns on the auxiliary atoms: either there are no auxiliary excitations (as in state $\vert{}ABC\rangle$), or there is exactly one excitation localized on atom $a$, $b$, or $c$. 
This structural property enables us to tailor the logical energy spectrum by selectively adjusting the detunings of these auxiliary atoms. 
In particular, the energy of the logical states $\vert{}Aa\rangle$, $\vert{}Bb\rangle$, and $\vert{}Cc\rangle$ can be independently shifted relative to the fourth reference state ($\vert{}ABC\rangle$) by adjusting the detunings of the auxiliary atoms. 
Therefore, the degeneracy of all four logical states can be restored by computing the vdW energy difference $\delta E_{m,n}$ between the reference state $\ket{m}=\ket{ABC}$ and each other logical state $\ket{n} \in \{\ket{Aa}, \ket{Bb}, \ket{Cc}\}$ and adjust the detuning on the corresponding auxiliary atom $a$, $b$, or $c$, to exactly mitigate $\delta E_{m, n}$.

In general, a module does not exist in isolation; rather, it has copy gadgets attached to its ports. 
In this setting, fixing the configuration on the module's ports uniquely determines the logical state across all connected copy gadgets via a well-defined $\mathbb{Z}_2$-continuation along the chains. 
This property allows us to define an extended set of logical states---in the example of the \textsf{3BODY} gadget: $\vert{}\overline{ABC}\rangle, \vert{}\overline{Aa}\rangle, \vert{}\overline{Bb}\rangle$, and $\vert{}\overline{Cc}\rangle$---that encompass both the module $g$ and its surrounding copy gadgets [see Fig.~\ref{fig:compensation}(d)].

When calculating the energy difference $\delta E_{m,n}$ from Eq.~\eqref{eq:VanishingTailsCondition}, where the indices $i$ and $j$ sum over all nodes within the module and its adjacent copy gadgets, the total sum accounts for three distinct contributions: intra-module interactions, cross-interactions between the module and its attached copy gadgets, and mutual interactions among the copy gadgets themselves. 
Crucially, all of these long-range contributions can be systematically aggregated and absorbed into the effective compensation weights on auxiliary atoms $a, b$, and $c$, because of the well-defined continuation of port excitations along copy gadgets.

For instance, let us select the logical states $\vert{}n\rangle = \vert{}\overline{Aa}\rangle$ and $\vert{}m\rangle = \vert{}\overline{ABC}\rangle$ and calculate the energy difference $\delta E_{m,n}$ via Eq.~\eqref{eq:VanishingTailsCondition}. In this case, $\delta E_{m,n}$ quantifies the net difference in vdW contributions between state $\vert{}n\rangle$ and state $\vert{}m\rangle$. 
A non-zero energy shift can be directly counteracted by setting the auxiliary compensation detuning on atom $a$ to $\Delta^{\rm comp}_a = -\delta E_{m,n}$, since this additional detuning selectively shifts state $\vert{}n\rangle$ while leaving state $\vert{}m\rangle$ and the other logical states unaffected. 
Analogously, compensation detunings for the remaining states, $\vert{}\overline{Bb}\rangle$ and $\vert{}\overline{Cc}\rangle$, are determined by taking $\vert{}\overline{ABC}\rangle$ as the universal reference state, whose energy remains invariant under local detunings on nodes $a, b$, or $c$.
With that, all four logical states on the module and connected copy gadgets can be made degenerate even under the presence of long-range vdW interactions.

The same method can be applied to all types of modules by introducing compensation weights $a, b$, and $c$ for each module, as depicted in Fig.~\ref{fig:homogenization}(a). 
As exemplified above, to account for a non-zero energy difference $\delta E_{m,n}$ in Eq.~\eqref{eq:VanishingTailsCondition}, these newly introduced compensation detunings yield an energy contribution of $-\delta E_g$ for each respective gadget $g$. 
These local contributions ensure that the global degeneracy requirement, $\delta E_{m,n} - \sum_{g} \delta E_g = 0$, is strictly satisfied across the full layout.
Explicitly, these local conditions can be formulated as:
\begin{align}
    \sum_{(i,j) \in N_g} V_\mathrm{vdW}(r_{ij})(n_i n_j - m_i m_j) & \nonumber \\
    - \sum_{i \in V_g} \Delta^{\rm comp}_i (n_i - m_i) &= 0,
\end{align}
where the set of pairs $N_g$ defines the local interaction neighborhood, $V_g$ denotes the set of atoms belonging to gadget $g$, and $\vert{}n\rangle, \vert{}m\rangle$ are two logical states. 
Depending on whether $g$ represents a module or a copy gadget, the interaction set $N_g$ is defined as follows.
\begin{itemize}
\item[(i)] Copy gadget: $N_g$ encompasses internal interactions within gadget $g$ only, i.e., pairs $(i,j)$ with $i, j \in V_g$.
\item[(ii)] Module: $N_g$ additionally includes cross-interactions between module nodes $i \in V_g$ and nodes of adjacent copy gadgets $j \in V_h$, as well as direct interactions between copy gadgets attached to different ports of the module, $i \in V_{h_1}$ and $j \in V_{h_2}$.
\end{itemize}
In summary, the introduction of compensation detunings $\Delta^{\rm comp}_i$ allows for a modular integration of long-range vdW effects. 
These additional detunings are required exclusively on the ports of copy gadgets and on the designated auxiliary nodes ($a, b, c$) of modules, as indicated in Fig.~\ref{fig:homogenization}.
Although these contributions originate on internal, non-port nodes, we demonstrate in the following section that they can be systematically shifted onto the gadget ports. 
This simultaneous homogenization of both bare detunings and vdW corrections renders the compensation scheme fully compatible with the global, anchor-based approach introduced in this work. 
Ultimately, we incorporate this compensation strategy as an intermediate step to determine the exact spatial coordinates of the anchor atoms in globally driven setups.

\section{Homogenization}
\label{app:Homogenization}
\begin{figure}[t]
\begin{centering}
\includegraphics[width = \columnwidth]{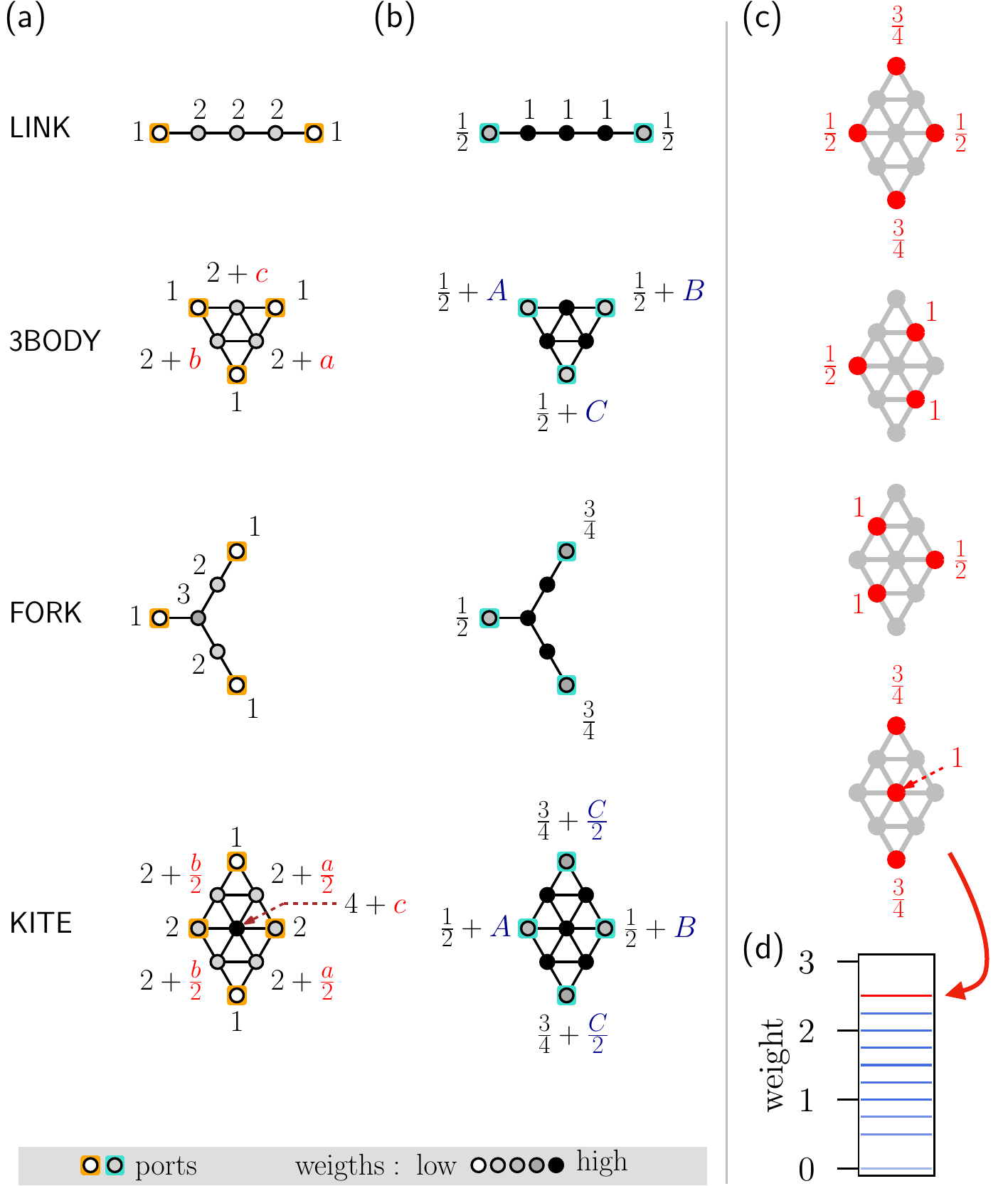}
\par\end{centering}
\protect\caption{\textit{Homogenization:} (a) Original Rydberg gadgets. 
Bare weights range from $1$ to $4$. 
Additional weights $a$, $b$, and $c$ include vdW compensation (see App.~\ref{app:LocalCompensationDetails}).
(b) Equivalent, homogenized gadgets have uniform weights of $1$ for all non-port nodes (atoms).
Note, that some of the `$1$' labels are omitted for clarity. 
Equivalence between the gadgets is understood as having the same solutions to the underlying \textsf{MWIS} problem. 
The additional weights $A, B$ and $C$ are related to $a, b$ and $c$ via Eq.~\eqref{eq:CompenstationHomo}.
(c) Solutions for the homogenized \textsf{KITE} gadget with weight spectrum (d) displaying the total weight of all possible independent sets.
}
\label{fig:homogenization}
\end{figure}
The set of gadgets introduced in Ref.~\cite{Lanthaler2023Ryblo} relies on local detunings, where in principle each atom could require its own distinct detuning $\Delta_i$. 
To illustrate this, let us focus on the \textsf{3BODY} gadget depicted in Fig.~\ref{fig:homogenization}. 
Here, the numbers adjacent to the nodes represent the \textsf{MWIS} weights, which are interpreted as local detunings by multiplying them by $\Delta$. 
Consequently, port atoms require a detuning of $\Delta$, whereas interior auxiliary atoms demand detunings of $(2+a)\Delta$, $(2+b)\Delta$, or $(2+c)\Delta$. 
Here, the baseline weight of $2\Delta$ stems from the \textsf{MWIS} gadget construction, while $a, b,$ and $c$ denote supplementary contributions from long-range vdW compensation [cf.~App.~\ref{app:LocalCompensationDetails}]. 
We now show that these vdW contributions can be systematically shifted onto the port atoms, thereby enabling our weak interaction approach using anchors positioned close to the ports.

For this purpose, we introduce a homogenized variant of the \textsf{3BODY} gadget, as shown in Fig.~\ref{fig:homogenization}(b). 
In contrast to the original construction, the homogenized \textsf{3BODY} gadget simplifies the interior to a uniform detuning of $\Delta$, while the detunings on the port atoms are modified to $(1/2 + A)\Delta$, $(1/2 + B)\Delta$, and $(1/2 + C)\Delta$. 
It is straightforward to check that if $A, B$, and $C$ satisfy
\begin{equation}
    \begin{bmatrix}
         A \\ B \\ C
    \end{bmatrix} =
    \frac{1}{2}
    \begin{bmatrix}
          1 & -1 & -1 \\ 
         -1 &  1 & -1 \\ 
         -1 & -1 &  1
    \end{bmatrix}
    \begin{bmatrix}
         a \\ b \\ c
    \end{bmatrix} ,
    \label{eq:CompenstationHomo}
\end{equation}
the homogenized gadget preserves the exact low-energy manifold defining the logical states: These lowest-energy configurations directly match those of the original gadget, and their corresponding energy eigenvalues are identical, up to a global constant shift of $-(a+b+c)\Delta/2$.
Hence, the original \textsf{3BODY} gadget can be replaced by its homogenized counterpart, as it faithfully implements the three-body parity constraint while absorbing the long-range van der Waals compensation shifts into the detunings of the ports.

Homogenized variants of other gadgets are displayed in Fig.~\ref{fig:homogenization}(b). 
The \textsf{LINK} gadget is trivial in this regard, as its internal detunings are already naturally homogenized and only a trivial change of scale has to be applied. 
Another useful copy gadget, which we denote as the \textsf{FORK} gadget, arises from the amalgamation of three \textsf{LINK}s; its homogenized version, however, cannot be directly obtained from the amalgamation of three homogenized link gadgets, but requires to adjust the local detuning pattern. 
The \textsf{KITE} gadget can also be homogenized such that local detunings are only required on its four ports, as shown in Fig.~\ref{fig:homogenization}(b). 
Furthermore, the compensation contributions $a$, $b$, $c$ can be transferred to ports using the exact same matrix transformation introduced in Eq.~\eqref{eq:CompenstationHomo}~\footnote{
In the case where the internal compensation weights $a, b,$ and $c$ vanish, there exists a unique homogenized version of the \textsf{KITE} gadget. 
In contrast, \textsf{FORK} gadgets admit a continuous family of homogenized parameter configurations. 
Specifically, if we denote the detuning weight on the left port as $l$ and on the right ports as $r$, the gadget remains homogenized and logically equivalent to the standard \textsf{FORK} as long as $l \in [0,1]$ and $r = (l+1)/2$. 
In this work, we select $l = 1/3$, as this choice maximizes the energy gap between the logical subspace and the high-energy spectrum.
}.
To verify the homogenized \textsf{KITE} gadget we draw the four logical states in Fig.~\ref{fig:homogenization}(c), where Rydberg excitated atoms are drawn in red. 
All configurations share an identical and maximal total weight (i.e., lowest energy) among all possible independent set states, as shown in Fig.~\ref{fig:homogenization}(d), confirming the gadget's functionality.

\section{Mitigation of radial distance fluctuations}
\label{app:zcomp}

\begin{figure}[t]
    \centering
    \includegraphics[width=1\linewidth]{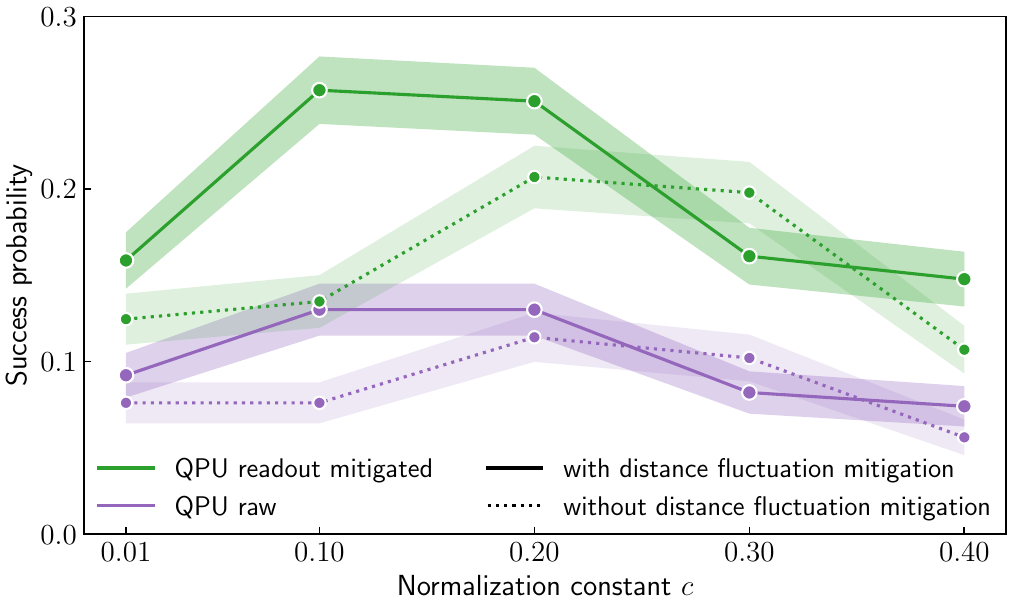}
    \caption{Comparison of experimental success probability for the \textsf{4BODY} gadget with (full lines) and without (dashed lines) mitigation of radial distance fluctuations.
    Raw experimental data is shown in purple, readout error mitigated data in green.
    For details see the caption of Fig.~\ref{fig:4body} in the main text.
    }
    \label{fig:sigma_z_compensation}
\end{figure}

A single atom trapped in an optical tweezer has a finite three-dimensional position distribution characterized by Gaussian standard deviations $(\sigma_x,\sigma_y,\sigma_z)$.
Because an optical tweezer is tightly focused, the position fluctuation is typically larger along the beam direction than orthogonal to it, i.e., $\sigma_z \gg \sigma_x,\sigma_y$.
For the performed experiments we use $\sigma_x=\sigma_y=0.14\,\mu$m and $\sigma_z=0.8\,\mu$m. 
Since the effective detunings $\dwl{}$ are programmed through direct interatomic interactions between anchors and other atoms, distance fluctuations can constitute a significant source of error.
While the position fluctuations of the individual atoms do not shift their mean positions, they do modify the average distance between two atoms.
Specifically, fluctuations radial to the connecting line between two atoms systematically increase their average distance, while longitudinal fluctuations yield no such systematic shift.
In the following, we describe how to mitigate this systematic distance increase from radial fluctuations.

Assume that two atoms are separated along the $x$ axis by a distance $R$. Under small relative displacements $\delta\mathbf{r} = (\delta x,\delta y,\delta z)$ satisfying $|\delta x|,|\delta y|,|\delta z| \ll R$, the actual distance becomes
\begin{align*}
R' &= \sqrt{(R+\delta x)^2+\delta y^2+\delta z^2}
\\
&\approx R+\delta x +\frac{\delta y^2+\delta z^2}{2R} + \cdots,
\end{align*}
where we applied a Taylor expansion in the second row.
Assuming independent Gaussian position fluctuations in each trap, we have $\langle \delta a \rangle = 0$ and $\langle\delta a^2 \rangle = 2\sigma_{a}^2$ for $a \in \{x, y, z\}$.
Since typical optical tweezer parameters satisfy $\sigma_y \ll \sigma_z$, the dominant contribution to the average distance becomes
\begin{equation}
    \langle R' \rangle \approx R  + \frac{\sigma_{z}^2}{R}. 
    \label{eq:distance_increase_from_fluctuations}
\end{equation}
Therefore, the average distance increases due to radial fluctuations.

We correct for this systematic distance increase between anchors and ports / \textsf{LINKs} during the anchor placement process.
In particular, when computing the anchor distance $\danch{e}$ via Eq.~\eqref{eq:weak_eqal_anchor} and Eq.~\eqref{eq:weak_eqal_link}, we replace 
\begin{equation}
    \Delta^{\rm weak}_{(\mathrm{logical}), e} (\danch{e}) \rightarrow 
    \Delta^{\rm weak}_{(\mathrm{logical}), e} (\tilde{d}_{\mathrm{anchor}, e}), 
\end{equation}
where $\tilde{d}_{\mathrm{anchor}, e} = \danch{e} + \sigma_z^2 / \danch{e}$ is computed according to Eq.~\eqref{eq:distance_increase_from_fluctuations}.

Figure~\ref{fig:sigma_z_compensation} illustrates the effect of mitigating radial distance fluctuations. 
Here, we present the mitigated experimental data for the \textsf{4BODY} gadget from Fig.~\ref{fig:4body}(e) of the main text (solid lines) alongside a new, unmitigated dataset where no distance corrections were applied during anchor placement (dashed lines). 
The applied mitigation generally yields higher success probabilities, confirming its effectiveness. While we observe a small reduction in success probability at $\pef=0.3$ for the mitigated case, the deviation remains within the experimental error.

\section{Readout error mitigation}
\label{app:readout_mitigation}

\begin{figure}[t]
    \centering
    \makebox[\linewidth]{\includegraphics[width=1\linewidth]{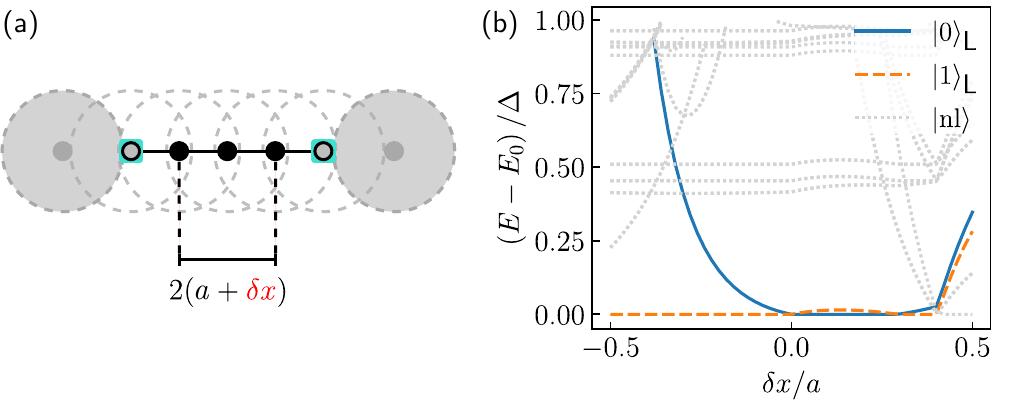}}
    \caption{Modifying \textsf{LINK} layout to induce an effective logical field $\dwl{}$.
    (a)
    We consider a \textsf{LINK} of length five with two anchor atoms inducing the required port detunings. Instead of placing an additional link anchor [cf.~Figs.~\ref{fig:weak_interaktion}(c, d)] the positions of the two atoms next to the center are shifted by $\delta x$. 
    (b)
    Low-energy spectrum as a function of $\delta x$. Logical states $\ket{0}_\textsf{L}$ ($\ket{1}_\textsf{L}$) are shown with blue solid (orange dashed) lines, non-logical states with gray dotted lines.
    }
    \label{fig:link_interatomic_distance_change}
\end{figure}

Readout errors during the final measurement can flip bits in the measured qubit state compared to the ideally measured state. They can therefore be treated as a mapping $T$ from the ideal bitstring probability distribution $P^{\rm ideal}$ to the observed distribution $P^{\rm measured}$:
\begin{equation}
  P^{\rm measured} = T P^{\rm ideal} \ .
  \label{eq:readout_error_mitigation}
\end{equation}
Assuming independent, uncorrelated readout errors for each atom, the mapping $T$ on $N$ qubits (atoms) can be written as 
\begin{equation}
T = 
\begin{pmatrix}
    1-p(1|0) & p(0|1) \\
    p(1|0) & 1-p(0|1)
\end{pmatrix}^{\otimes N}
\end{equation}
where $p(1|0)$ denotes the false positive error rate (i.e., measuring a Rydberg state instead of the actual ground state) and $p(0|1)$ the false negative error rate (i.e., measuring a ground state instead of the actual Rydberg state). Throughout the manuscript, we use the independently characterized values $p(1|0)=0.02$ and $p(0|1)=0.10$.

In principle, the ideal bitstring distribution can be recovered by solving Eq.~\eqref{eq:readout_error_mitigation} for $P^{\rm ideal}$. In practice, several challenges arise: directly inverting $T$ is generally unstable, making alternative linear algebra solvers necessary; 
$T$ is a $2^N \times 2^N$ matrix, rendering full storage impossible for large qubit numbers; 
and Eq.~\eqref{eq:readout_error_mitigation} operates on the full probability distribution, meaning mitigation relies heavily on a high sampling rate to overcome shot noise---a difficult requirement to satisfy for large $N$.

To mitigate readout errors efficiently, we employ the ``matrix-free measurement mitigation'' (M3) method~\cite{Nation2021}. This approach operates exclusively within the reduced subspace of observed bitstrings $P^{\rm measured}$ and applies an iterative solver to estimate $P^{\rm ideal}$.

In Figs.~\ref{fig:4body}, \ref{fig:kite_experiment}, and \ref{fig:4body_link_length}, the positive impact of this readout error mitigation is clearly demonstrated. However, for the larger experiment on the LHZ layout in Fig.~\ref{fig:lhz_experiment}, the low number of measured samples relative to the vast Hilbert space size, combined with the method's restriction to only the measured bitstrings, renders the mitigation ineffective for that specific dataset.

\hspace{1cm}
\section{Weak interactions without anchors}
\label{app:weak_without_anchors}
In the main text, we focused solely on weak interactions from additionally placed anchor atoms, leaving the basic gadget geometries proposed in Ref.~\cite{Lanthaler2023Ryblo} unaltered. 
However, slightly modifying the positions of atoms within the gadget also induces weak interactions on nearby atoms, generating effective local detunings and enabling more atom-efficient geometries.

Here, we demonstrate this approach for the logical detuning $\dwl{}$ on a \textsf{LINK}. 
Instead of placing a link anchor to change $\dwl{}$ based on its distance, we symmetrically shift the positions of the atoms adjacent to the center atom within the \textsf{LINK}, as shown in Fig.~\ref{fig:link_interatomic_distance_change}(a). 
Provided the position modification $\delta x$ is sufficiently small, the Rydberg blockade pattern remains intact, inducing only additional \textit{weak interactions}. Figure~\ref{fig:link_interatomic_distance_change}(b) shows the resulting low-energy spectrum as a function of $\delta x$. 
Indeed, for sufficiently small $|\delta x|$, the two logical states $\ket{0}_\textsf{L}$ (blue solid line) and $\ket{1}_\textsf{L}$ (orange dashed line) remain well separated from non-logical states (gray dotted lines), confirming that the Rydberg blockade picture is preserved in this regime.
However, particularly when moving the two atoms closer together ($\delta x < 0$), we observe a splitting between the logical states that scales with $|\delta x|$. As described in the main text [cf.\ Sec.~\ref{sec:weak_interactions} and Fig.~\ref{fig:weak_interaktion}(c, d)], this is equivalent to an effective logical detuning $\dwl{}(\delta x) > 0$.

Conversely, when the distance between these atoms is increased ($\delta x > 0$), the energy difference between the logical states remains minimal, inducing only a tiny $\dwl{} < 0$. Note that this asymmetry is a result of the specific \textsf{LINK} setup chosen here.

Finally, if $|\delta x|$ becomes too large, the blockade pattern breaks down, leading to a rapid proliferation of non-logical states. For the setup chosen here, this breakdown occurs at approximately $|\delta x| \approx 0.3 d$.

This example demonstrates that weak interactions can be tuned directly within gadgets by altering their internal structure, and that appropriate modification patterns can successfully induce desired local fields.

\end{document}